\documentclass[twocolumn,apj,numberedappendix,iop]{openjournal}
\usepackage[T1]{fontenc}
\usepackage{graphicx}	% Including figure files
\usepackage{amsmath}	% Advanced maths commands
\usepackage{newtxtext,newtxmath}    % PUT THIS ONE BACK IN IN OVERLEAF
\usepackage[dvipsnames]{xcolor}  
\usepackage[colorlinks,linkcolor=blue,citecolor=blue,urlcolor=blue ]{hyperref}
\usepackage{amssymb}	% Extra maths symbols
\usepackage{dsfont}
\usepackage{ulem}
\usepackage{soul}
\usepackage{orcidlink}
\usepackage{hyperref}

\usepackage{natbib}
\usepackage{fontawesome}
\usepackage{lineno}

\usepackage{macros_vdb} % my personal macros

\begin{document}

%%% uncomment this if you want linenumbers
%%\linenumbers

\title{BASILISK III. Stress-testing the Conditional Luminosity Function model}

\shorttitle{BASILISK III. Stress-testing the CLF model}
\shortauthors{Mitra and van den Bosch}

\author{Kaustav Mitra\orcidlink{0000-0001-8073-4554}}
\author{Frank C. van den Bosch\orcidlink{0000-0003-3236-2068}}   

\affiliation{Department of Astronomy, Yale University, PO. Box 208101, New Haven, CT 06520-8101}

\email{kaustav.mitra@yale.edu}

\label{firstpage}

%%%%%%%%%%%%%%%%%%%%%%%%%%%%%%%%%%%%%%%%%%%%%%%%%%%%%%%%%%%%%%%%%

\begin{abstract}
  The Conditional Luminosity Function (CLF) is an effective and flexible way of characterizing the galaxy-halo connection.  However, it is subject to a particular choice for its parametrization, which acts as a prior assumption. Most studies have been restricted to what has become a standard CLF parametrization with little to no variation. The goal of this paper is to investigate whether this model is sufficient to fully characterize the small-scale data extracted from spectroscopic surveys and to gauge how adding or removing degrees of freedom impact the inference regarding the galaxy-halo connection. After extensive validation with realistic mock data, we use \Basilisc, a highly constraining Bayesian hierarchical tool to model the kinematics and abundance of satellite galaxies, to test the standard CLF model against a slew of more flexible variants. In particular, we test whether the SDSS data favour any of these variants in terms of a goodness-of-fit improvement, and identify the models that are sufficiently flexible, beyond which additional model freedom is not demanded by the data. We show that some of these additional degrees of freedom, which have hitherto not been considered, result in a drastic improvement of the fit and cause significant changes in the inferred galaxy-halo connection. This highlights that an empirical model comes with an implicit prior about the parametrization form, which needs to be addressed to ensure that it is sufficiently flexible to capture the complexity of the data and to safeguard against a biased inference.
\end{abstract} 

%%%%%%%%%%%%%%%%%%%%%%%%%%%%%%%%%%%%%%%%%%%%%%%%%%%%%%%%%%%%%%%%%%%%


%%%%%%%%%%%%%%%%%%%%%%%%%%%%%%%%%%%%%%%%%%%%%%%%%%%%%%%%%%%%%%%%%%%%

\section{Introduction}
\label{sec:intro}

Galaxy formation in the $\Lambda {\rm CDM}$ cosmological framework hinges on the key idea that dark matter forms the large-scale structure, collapsing into virialized halos, and that gas falls into those gravitational potential wells, cools, and forms galaxies, which then grow, merge, and evolve in a hierarchical way. A data-driven approach to derive meaningful insights regarding the physics of galaxy formation is to constrain the galaxy-halo connection, or the statistical link between galaxy properties and that of their host halos, using observable statistics. It is usually modeled using either the halo occupation distribution (HOD) or the  conditional luminosity function (CLF). The former parametrizes the statistical link in terms of the mean number of galaxies of a given property (typically above some luminosity or stellar mass threshold) that occupy a halo of a given mass \citep[see e.g.][]{Zheng.etal.05, Zheng.etal.07, Zehavi.etal.11}, while the latter parametrizes the full luminosity (or stellar mass) distribution of galaxies in halos of given mass \citep[see e.g.][]{Yang.etal.04,vdBosch.etal.07, Cacciato.etal.09}. Note that these two characterizations of the galaxy-halo connection modeling are related to each other; integrating the CLF above some luminosity threshold yields the corresponding HOD model.

Clustering of galaxies \citep{Zehavi.etal.11} and weak gravitational lensing of background sources \citep{Mandelbaum.etal.06} are the two most common observables used to statistically constrain the halo masses of galaxies. Over the last two decades, numerous independent data-driven constraints have led to a broad consensus regarding the median relation between stellar mass and host halo mass (SHMR) of central galaxies \citep{Behroozi.etal.19}. It is now a cornerstone of galaxy formation theory \citep[see][for details]{Wechsler.Tinker.18}. Both semi-analytical models \citep{Croton.etal.16, Stevens.etal.2016} and hydrodynamical simulations \citep{McAlpine.etal.16, Pillepich.etal.2018a} of galaxy formation are often tested or calibrated against this SHMR. Interestingly, while there is consensus on the median relation, the level of intrinsic scatter in the SHMR is less well established. Most empirical studies that constrain the galaxy-halo connection using clustering, lensing, or other probes infer fairly low values for the scatter, in the range of 0.15 to 0.2 dex \citep{Yang.etal.08, More.etal.09b, Hearin.Watson.13, Cacciato.etal.13, Lange.etal.19b}. Although these are in reasonable agreement with the predictions from hydrodynamical simulations, they are significantly lower than the predictions from most semi-analytical models \citep[][]{Wechsler.Tinker.18, Porras-Valverde.etal.2023}. A notable outlier is the empirical study by \citet{Behroozi.etal.19}, which inferred a significantly higher scatter that is more in line with the semi-analytical models. In order to make progress, a more stringent and robust inference of the galaxy-halo connection is needed. 

Models for the galaxy-halo connection are a powerful tool for mapping what we can see in galaxy surveys to the underlying dark matter distribution that governs structure formation and dynamics in the Universe. Although such mapping, when restricted to large scales, can rely on bias expansion \citep[see][for a review]{Desjacques.etal.2018}, those techniques fail to capture the complexity at the quasilinear and smaller scales of survey data. Thus, over the last couple of decades, HOD/CLF modeling of the galaxy-halo connection has become an indispensable tool in using galaxy surveys to probe and constrain the matter distribution in the Universe and thus constrain cosmology \citep{Lange.etal.19c, Lange.etal.23}. Such analyses, mostly using galaxy clustering or weak lensing as the observables, often infer weaker matter clustering \citep{Leauthaud.etal.17} compared to what one predicts based on the cosmological inference from the cosmic microwave background (CMB) data obtained by the \Planc satellite \citep{Planck.18}. In terms of the cosmological parameters, this disagreement shows up as a tension in the 2D plane of $\Omega_\rmm$ (the mean matter density), and $\sigma_8$ (the amplitude of the linear matter power spectrum at the $8h^{-1}{\rm Mpc}$ scale) \citep[see][for a review]{Perivolaropoulos.&.Skara.2022}. The tension is strongest in the direction of a composite parameter $S_8 \equiv \sigma_8 (\Omega_\rmm/0.3)^{0.5}$, and is therefore generally called the $S_8$ tension \citep{DES_18a, Joudaki.etal.2020}.

The persistence of the $S_8$ tension over the last decade has led to speculations about beyond-standard model physics to resolve it \citep[see][for a review]{Abdalla.etal.2022}. However, it could also be a result of systematics in low-redshift analyses. For example, clustering of halos is well known to depend not only on their mass but also on their formation history, a phenomenon known as halo assembly bias \citep{Gao.etal.05, Gao.White.07, Dalal.etal.08, Li.etal.2008}. It could affect both galaxy clustering and weak lensing studies \citep{Zentner.etal.14, Beltz-Mohrmann.etal.2020}. However, it is often not taken into account, which can lead to biases in galaxy-halo connection inference and cosmological analyses. Another key challenge to modeling the small-scale data from galaxy surveys arises from baryonic effects. The radiation-hydrodynamics related to galaxy formation physics can impact the matter distribution especially on small scales \citep{Chisari.etal.2019, Amodeo.etal.2021, McCarthy.etal.2024, Baggen.etal.25}, something that is often ignored in cosmological analyses. 

A third source of systematics, and the focus of this paper, is the galaxy-halo connection model itself, specifically the hidden priors that are introduced by assuming a functional form for such empirical models. Typically, the models used to parameterize the HOD are rather limited, simply assuming a power-law scaling between the number of satellites and halo mass and an error function to quantify the expectation value for the number of centrals as a function of halo mass \citep[][]{Berlind.etal.03, Zheng.etal.05, Zehavi.etal.11, Wang.etal.22, Guo.etal.10}. Only a few studies have probed the limitations of this model and examined extensions or enhancements using a data-driven approach. \citet{Sinha.etal.17} used multiple observables derived from the seventh data release of the Sloan Digital Sky Survey (SDSS DR7) \citep{Abazajian.etal.09}, including galaxy clustering, galaxy number density, and group multiplicity, to statistically test the robustness of the standard HOD model. Although they find the model to be sufficiently flexible to accurately fit each set of summary statistics separately, it fails to achieve a joint inference. Although this could be interpreted as a cosmological tension, the authors argue that it might also indicate that the parametrization of the HOD model is too restrictive. \citet{Beltz-Mohrmann.etal.2020} tested the validity of the standard HOD model by applying it to the Illustris \citep[][]{Genel.etal.14} and EAGLE \citep[][]{Schaye.etal.15} simulations and reached a similar conclusion that the standard HOD model is too restrictive to accurately reproduce the galaxy-halo connection of simulated galaxies.

Unlike the HOD, the parametrization of the CLF has undergone some modifications since its introduction in \citet{Yang.etal.03} and \citet{vdBosch.etal.03}. In particular, the earliest CLF models did not distinguish between central and satellite galaxies and assumed that the two are drawn from the same underlying distribution, the central being simply the brightest galaxy in its halo \citep[see e.g.][]{vdBosch.etal.03b, vdBosch.etal.05c, vdBosch.etal.07, Yang.etal.04}. However, galaxy formation theories dictate that central galaxies are the repositories of cooling flows and are expected to grow in mass by cannibalizing their satellites, whereas satellite galaxies are subjected to a number of environmental processes that impact their mass and star formation rate, such as tidal stripping, ram pressure stripping, and strangulation \citep[][]{MBW10}. Hence, central galaxies are expected to be rather distinct from satellites. Indeed, several studies have shown that centrals and satellites have distinct properties \citep[e.g.,][]{Weinmann.etal.06, vdBosch.etal.08a} and that they are distinct compared with predictions based on extreme value statistics \citep[e.g.,][]{Lin.etal.10, Shen.etal.14}\footnote{Though there are also claims to the contrary \citep[e.g.,][]{Dobos.Csabai.11, Paranjape.Sheth.12, WangE.etal.18}.}. \citet{Cooray.Milosavljevic.05} and \citet{Cooray.06} were the first to split the CLF into distinct central and satellite components. In particular, they modeled the central CLF as a log-normal distribution and the satellite CLF as a power-law distribution with a cutoff at the bright end. Subsequently, in their analysis of the occupation statistics of galaxy groups in the \citet{Yang.etal.07} SDSS group catalog, \citet{Yang.etal.08} adopted a similar form, replacing the satellite CLF with a modified Schechter function. This form had previously been advocated by \citet{Zheng.etal.05} based on an analysis of occupation statistics in semi-analytical models of galaxy formation and has become the standard functional form adopted in virtually all CLF-based studies since \citep[][]{Moster.etal.10, More.etal.11, Leauthaud.etal.11, Yang.etal.12, vdBosch.etal.13, Cacciato.etal.09, Cacciato.etal.13, Shi.etal.16, Lange.etal.18, Lange.etal.19b}. 

Somewhat surprisingly, although the CLF is more flexible than its HOD counterpart, to our knowledge, no study has ever examined how restrictive the CLF parametrization actually is and whether additional degrees of freedom impact the inference. The goal of this paper is to address this issue by comparing the inferences on the galaxy-halo connection obtained using data from the SDSS DR7 using CLF parametrizations with different degrees of freedom. Specifically, we use satellite kinematics, satellite abundances, and the total luminosity function of galaxies, as modeled by \Basilisk \citep[][]{vdBosch.etal.19, Mitra.etal.24}, to test the standard CLF parametrization. \Basilisk is a novel Bayesian hierarchical tool that primarily forward models the projected phase-space distribution of galaxies around potential central galaxies identified in a redshift survey, while accounting for incompleteness, interlopers, impurities, survey systematics, and selection effects. Satellite kinematics, an under-utilized technique to infer the galaxy-halo connection, is based on the fact that the spatial and velocity distributions of satellites carry information about the halo potential well in which they orbit \citep[][]{Brainerd.Specian.03, Prada.etal.03, vdBosch.etal.04, Norberg.etal.08, More.etal.09b, More.etal.11}. \citet{Lange.etal.19a, Lange.etal.19b} transformed this technique into a precision tool to constrain the galaxy-halo connection on par with clustering and lensing analyses. \citet{vdBosch.etal.19} (hereafter Paper~I) introduced \Basilisk as a proof of concept, while \citet{Mitra.etal.24} (hereafter Paper~II) significantly improved the machinery with sophisticated modeling of systematics to turn \Basilisk into a robust and unbiased tool to infer the galaxy-halo connection with extreme precision, ensuring full and efficient extraction of information from small-scale data. Paper~II also applied \Basilisk to the SDSS DR7 data to constrain the CLF of central and satellite galaxies, following a fairly standard parametrization. Here, we repeat the same analysis, but using various extensions and modifications of the CLF parametrization. Our goal is twofold: (i) to examine how hidden priors related to the functional form of the galaxy-halo model impact inference on the galaxy-halo connection, and (ii) to construct a more flexible and accurate data-driven model of the galaxy-halo connection.

This paper is organized as follows. Section~\ref{sec:data} describes the dataset that we use for our analysis. Section~\ref{sec:modeling} briefly summarizes how \Basilisk uses these data to constrain the galaxy-halo connection, and Section~\ref{sec:GHC_models} introduces the different CLF parametrizations used in this study. In Section~\ref{sec:mock_validation} we apply \Basilisk with the different CLF parametrizations to mock data, demonstrating unbiased recovery of the galaxy-halo connection regardless of the model freedom assumed. In Section~\ref{sec:result} we apply the same analysis to the SDSS DR7, which contrary to the mock data yields results that strongly depend on the CLF parametrization assumed. We elaborate on these findings in Section~\ref{sec:discussion} where we discuss the implications in terms of more directly observable quantities. Finally, Section~\ref{sec:concl} summarizes our findings.

Throughout this paper, we assume a flat $\Lambda{\rm CDM}$ cosmology with \Planc best-fit parameters $\Omega_{\rm m}=0.315$, $\sigma_8 = 0.811$, spectral index $n_{\rm s}=0.9649$, Hubble parameter $h = (H_0/100 \kmsmpc) = 0.6736$, and baryon density $\Omega_{\rm b} h^2 = 0.02237$, which are the best-fit values corresponding to \texttt{Plik} likelihood of \Planc TT+TE+EE+lowE+lensing \citep{Planck.18}.
\begin{figure*}
\centering
\includegraphics[width=0.9\textwidth]{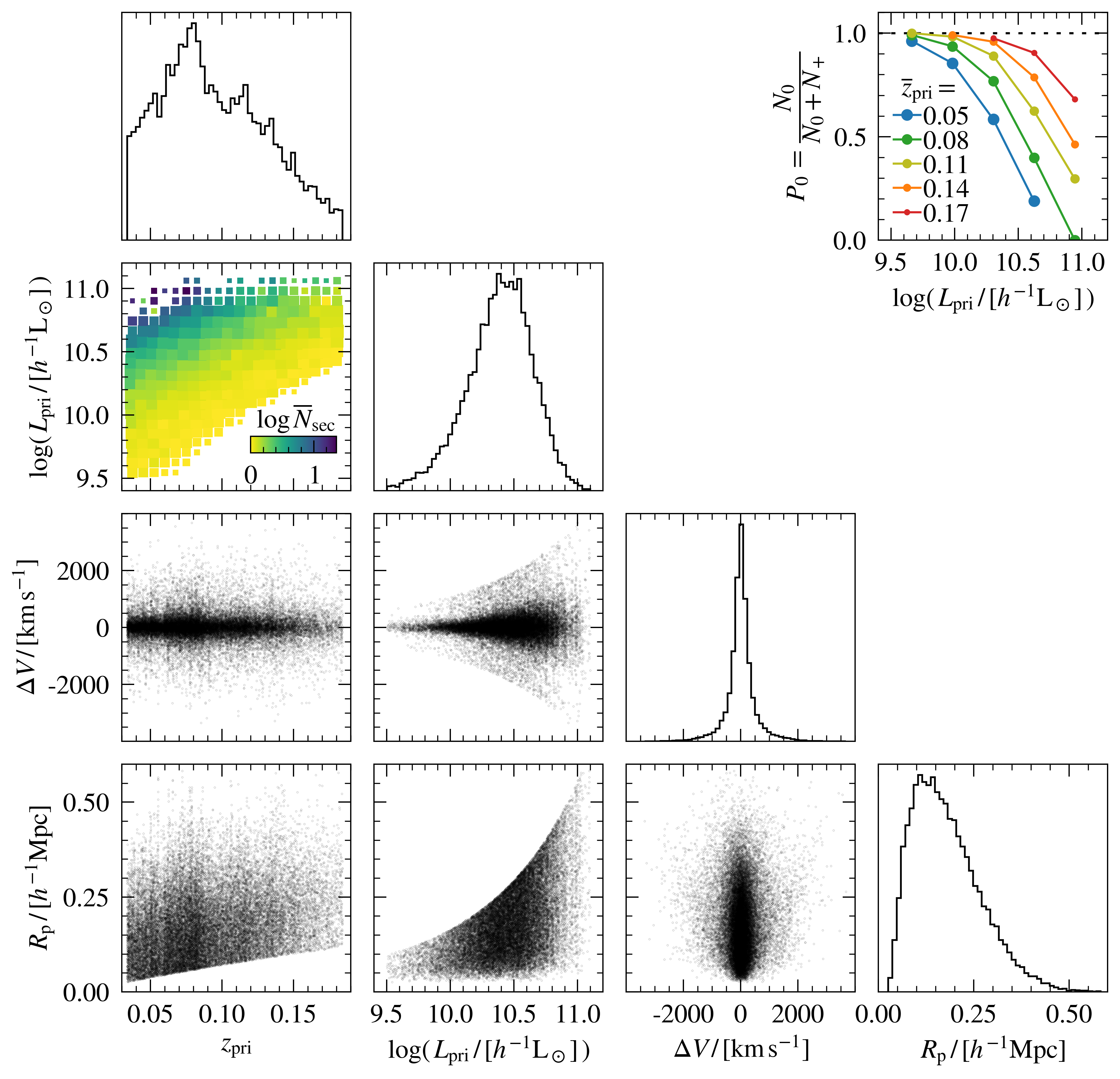}
\caption{The panels in the lower-left triangle comprise the complete depiction of the satellite kinematics data, along with their mutual dependencies. The histograms along the diagonal indicate, from top-left to bottom-right, the distributions of the redshift of the primaries ($\zpri$), the primary luminsoties ($\Lpri$), the line-of-sight velocity differences of primary-secondary pairs ($\Delta V$), and their projected separations ($R_{\rm p}$). The $(\zpri,\Lpri)$ co-dependence is the form of a 2D binned histogram where the bins are color-coded according to the mean number of secondaries in each bin. The other co-dependencies are in the forms of scatter plots, discussed further in the text. The most illuminating is the $(\Lpri, \Delta V)$ scatter plot, clearly showing that the velocity dispersion is higher for more luminous primaries, indicating that they reside in more massive halos. The panel in the top-right corner shows the complementary information that is not captured in the satellite kinematics data. Here, each line represents the fraction of all primaries in a redshift bin that has zero detected secondaries, as a function of the primary galaxy luminosity. This information is contained in the satellite abundance data vector, $\bD_{\rm Ns}$, which is also modelled in \Basilisc's framework.}
\label{fig:sdss_data}
\end{figure*}

\section{Data}
\label{sec:data}

As in \papII, we use data from the Sloan Digital Sky Survey (SDSS) seventh data release (DR7) \citep{Abazajian.etal.09}. \Basilisc's modeling of satellite kinematics and satellite abundance data requires that we first identify central and satellite galaxies from a redshift survey. This identification is inherently flawed. For example, sometimes bright satellites are selected as potential centrals; we refer to those as impurities. Similarly, some of the galaxies that are selected as potential satellites around a central do not truly belong to the same host halo; these are called interlopers. Due to these misidentifications, we reserve the words ``centrals'' and ``satellites'' for the true central and satellite galaxies, and we refer to the potential centrals and satellites selected from the survey as ``primaries'' and ``secondaries''.

The selection of primary and secondary galaxies is essentially based on an isolation criterion. A galaxy at redshift $z$ is selected as a primary if it is the most luminous one in a conical volume, defined by the opening angle $\theta_{\rm ap}^{\rm pri} \equiv \Rh / d_{\rm A}(z)$ and the redshift range $z \pm (\Delta z)^{\rm pri}$, centered around that galaxy. Any galaxy is disqualified from being a primary if it falls within the primary selection cone of another more luminous galaxy. Here $d_{\rm A}(z)$ is the angular diameter distance to redshift $z$, while the parameters $\Rh$ and $(\Delta z)^{\rm pri} \equiv \dVh/c (1+\zpri)$ both scale with the luminosity of the galaxy on which the selection cone is centered. After all primaries have been selected from the survey, the fainter galaxies around each primary falling inside a similar secondary selection cone, defined by parameters $\Rs$ and $\dVs$, are selected as the associated secondaries. Note that \Basilisc's results are highly insensitive to the exact details of the selection process, because all selection effects are forward modeled in \Basilisc's Bayesian framework.

The selection criteria for primaries and secondaries used in this work are identical to those in \papII, and thus the resultant samples of primary and secondary galaxies are also the same. With a luminosity and redshift cut of $9.504 \leq \log[\Lpri/(h^{-2}\Lsun)] \leq 11.104$ and $0.034 \leq z \leq 0.184$, we have a total of $\sim 165,000$ primaries. Among those, only $N_+=18,373$ primaries have at least one secondary. The total number of secondaries, and thus the total number of primary-secondary pairs for which kinematic data is available, is $30,431$. For each primary-secondary pair, we compute the projected separation
\begin{equation}\label{Rpdef}
\Rp = d_\rmA(\zpri) \, \vartheta\,,
\end{equation}
and their line-of-sight velocity difference,
\begin{equation}\label{dvdef}
\dV = c \, \frac{(\zsec - \zpri)}{1 + \zpri}\,.
\end{equation}
Here, $\zpri$ and $\zsec$ are the observed redshifts of the primary and secondary, respectively, $c$ is the speed of light, and $\vartheta$ is the angular separation between the primary and secondary in the sky. 

The full data set is organized into two separate data vectors. The first is the satelite kinematics (SK) vector,
\begin{equation}\label{eqn:SK_data}
\bD_{\rm SK} = \bigcup\limits_{i=1}^{N_{+}} \, \left( \{\dVij, \Rij | j=1,...,\Nsi \} | \Lci, \zci, \Nsi \right)\,,
\end{equation}
where $\Nsi$ is the number of secondaries associated with primary $i$ and the union is over all $N_{+}$ primaries with at least one secondary. Note that $\Lci$, $\zci$, and $\Nsi$ are treated as {\it conditionals} for the data $\{\dVij, \Rij | j=1,...,\Nsi \}$ (see \paperII for a justification of this approach).

The second data vector is given by
\begin{equation}\label{eqn:Ns_data}
\bD_{\rm Ns} = \bigcup\limits_{i=1}^{N_{\rm Ns}} \, \left( \Nsi \, | \, \Lci, \zci\right)\,,
\end{equation}
and characterizes the number of secondaries for a given primary (with luminosity $\Lci$ at redshift $\zci$). Here, the union is over a subset of $N_{\rm Ns} \simeq \calO(N_{+})$ primaries from the entire sample of primaries, independent of how many secondaries they have (i.e., including ``lonely primaries'' with zero secondaries). The reason for using a random subset rather than all $\Npri$ primaries is to limit the computational cost (see \paperII for details). Note that traditional studies of satellite kinematics typically exclude lonely primaries from their analysis. \Basilisc, however, uses this vast amount of data as it holds crucial information regarding the galaxy-halo connection. As discussed in Section~\ref{sec:modeling} below, \Basilisk forward-models the full $\bD_{\rm SK}$ and $\bD_{\rm Ns}$ distributions, given a set of galaxy-halo connection parameters, $\btheta$, and an assumed cosmology. 

Fig.~\ref{fig:sdss_data} showcases the full multi-dimensional data that \Basilisk uses as input. The histograms along the diagonal show the distributions of primary galaxy redshift, luminosity, the line-of-sight velocity difference of primary-secondary pairs, and their projected separations, all taken from the $\bD_{\rm SK}$ sample of the SDSS DR7 data. The off-diagonal panels in the lower left triangle show the mutual scatter plots among any two of these observables, highlighting their complex mutual co-dependence. The $(\Lpri,\zpri$) panel shows a binned histogram instead of a scatter plot, where each bin is represented by a square whose size depends on the number of points in the bin, and the color of the square indicates the mean number of secondaries per primary.

The sharp edge in the $(R_{\rm p},\zpri)$ panel of Fig.~\ref{fig:sdss_data} corresponds to the effect of fiber collisions. All secondaries with $R_{\rm p} / d_{\rm A}(\zpri) < \theta_{\rm cut}$ are excluded to mitigate the effect of missing redshifts within the fiber placement radius of the survey, which in the case of SDSS is $\theta_{\rm cut} = 55$ arcseconds. The sharp edge in the $(R_{\rm pri},\Lpri)$ panel corresponds to the fact that the seondary selection cone has a maximum radius that scales quadratically with $\log \Lpri$, to account for the fact that more luminous primaries typically reside in larger halos (see \paperII for details). Similarly, the sharp edges in the $(\Delta V, \Lpri)$ panel reflect luminosity-dependent secondary selection, i.e., the fact that $\dVs$ of the selection cone scales quadratically with $\log \Lpri$ to account for higher velocity dispersion in more massive halos. But even without these sharp edges, the fact that the velocity dispersion increases with $\Lpri$ is clearly evident from the triangular shape of the dense distribution of points in this panel. This indicates that more luminous galaxies must be residing in more massive halos. Finally, the $(R_{\rm p},\Delta V)$ panel shows the 2D projected phase-space distribution of points. This is the maximum dynamical information that can be obtained from a galaxy redshift survey, and this is what \Basilisk forward models to constrain the galaxy-halo connection.

The top right panel of Fig.~\ref{fig:sdss_data} shows complementary information regarding the occupation statistics of satellite galaxies. It shows the probability that a primary has zero associated secondaries, $P_0 \equiv N_0/(N_0+N_+)$, as a function of primary luminosity and for five different redshift intervals. Here, $N_0$ and $N_+$ are the numbers of primaries with zero secondaries and at least one secondary, respectively. As expected, this probability decreases as a function of $\Lpri$, because (a) more luminous primaries live in more massive halos that have a larger number of secondaries, and (b) a higher $\Lpri$ corresponds to a larger secondary selection volume. For any given primary luminosity, the probability $P_0$ is higher for larger $\zpri$ because the luminosity threshold corresponding to the flux limit of the survey increases with redshift, or, in other words, because fainter satellites become spectroscopically undetectable as one goes to higher redshifts. \Basilisk fully utilizes this information, typically ignored in traditional analyses of satellite kinematics, in the form of the data vector $\bD_{\rm Ns}$.

Note that $\bD_{\rm SK}$ and $\bD_{\rm Ns}$ capture information on the projected phase-space distribution and abundance of the secondaries {\it given} the luminosity and redshift of the primary. \Basilisk does not forward model the distribution of $(\Lpri, \zpri)$, but instead utilizes that information by simultaneously fitting the total luminosity function of all galaxies. As in \papII, we compute the number densities of galaxies, $n_{\rm gal}$, in $N_{\rm LF}$ luminosity bins, which yields the complementary data vector 
\begin{equation}\label{eqn:LF_data}
\bD_{\rm LF} = \bigcup\limits_{k=1}^{N_{\rm LF}} \, n_{\rm gal}(L_{k,{\rm min}}, L_{k,{\rm max}})\,.
\end{equation}
Here, the luminosities inside the brackets represent the minimum and maximum limits of the $k^{\rm th}$ luminosity bin, with $L_{k+1,{\rm min}} = L_{k,{\rm max}}$. This ensures that the luminosity bins are non-overlapping and exhaustive. We follow \cite{Lange.etal.19a, Lange.etal.19b} and Paper~II and use the number density of galaxies in ten, 0.15~dex bins in luminosity, ranging from $10^{9.5}$ to $10^{11} \Lsunh$. These are computed using the corresponding volume-limited subsamples, carefully accounting for the SDSS DR-7 footprint. 

\subsection{Mock Data}
\label{sec:mockdata}

To test and validate \Basilisc's machinery, we first apply it on realistic SDSS-like mock data. This mock survey is produced by populating halos and subhalos in the SMDPL simulation \citep {Klypin.etal.16} with central and satellite galaxies following a fiducial conditional luminosity function (CLF). The galaxies are given the phase-space coordinates of the 10 percent most bound particles in each subhalo, and thus no assumption is made about positional or velocity isotropy or about whether the satellites are in a state of dynamical equilibrium in the halo. A virtual observer is placed within the simulation volume to create a mock galaxy redshift survey, where each galaxy redshift includes the cosmological redshift based on the galaxy's comoving distance in the simulation box, the redshift due to its peculiar velocity along the line-of-sight, and a redshift error of $\sigma_v = 15 \, {\rm km}/{\rm s}$ comparable to that in the SDSS DR7 data. The SDSS flux limit and footprint are applied to the data to account for survey selection, aperture incompleteness, spectroscopic incompleteness, and fiber collisions. See Papers~I and~II for more details on the creation of the mock data.

Once the mock survey is obtained with all systematics representative of the true SDSS data, we apply the primary and secondary selection criteria as described above. \paperII optimized the selection process to minimize the effects of impurities, and we use the same selection cone parametrization as in that paper, for both the mock data and the real SDSS data.

\section{Modeling the Data}
\label{sec:modeling}

As explained in the previous section, the data are parsed into three complementary data vectors $\bD_{\rm SK}$, $\bD_{\rm Ns}$, and $\bD_{\rm LF}$. \Basilisk maximizes the corresponding total logarithmic likelihood
\begin{align}\label{eqn:likelihood}
\ln \calL &= \ln \big[ P(\bD_{\rm SK}| \btheta) \times P(\bD_{\rm Ns}| \btheta) \times P(\bD_{\rm LF}| \btheta) \big] \nonumber
\\
&= \ln \calL_{\rm SK} +  \ln \calL_{\rm Ns} + \ln \calL_{\rm LF}\,,
\end{align}
where $\btheta$ is the vector of model parameters, which includes all the galaxy-halo connection parameters, three nuisance parameters used to model interlopers, and one parameter specifying the average velocity anisotropy of satellite galaxies. A detailed specification of the various model parameters is given in Section~\ref{sec:GHC_models}.

Although it appears that in equation~(\ref{eqn:likelihood}), the likelihood of each component is treated as an independent probability, in reality we can write it this way only because $\bD_{\rm SK}$ holds the information in $\bD_{\rm Ns}$ and $\bD_{\rm LF}$ as conditionals, and similarly $\bD_{\rm Ns}$ holds the information in $\bD_{\rm LF}$ as conditional. Thus, the mutual codependencies of the data are explicitly accounted for. 

In the following subsections, we briefly sketch how \Basilisk computes the three different likelihoods. Since a detailed description can be found in \papII, we focus only on the salient points, without going into detail. 

\subsection{Satellite Kinematics modeling}
\label{ssec:SK_model}

The satellite kinematics likelihood is computed using
\begin{align}
    \calL_{\rm SK} &= P(\bD_{\rm SK}| \btheta)  \nonumber
    \\
    &= \prod\limits_{i=1}^{N_{+}} \, \prod\limits_{j=1}^{\Nsi} P\left( \dVij, \Rij  \, | \, \Lci, \zci,  \Nsi \right)\,,
\end{align}
which makes the key assumption that the phase-space distribution of different primary-secondary pairs are independent of each other. Another assumption that is used throughout is that satellites are virialized steady-state tracers of the dark matter potential, which is assumed to be spherical with an NFW \citep[][]{Navarro.etal.97} density profile and a concentration given by the concentration-mass relation of \citet{Diemer.Kravtsov.15} with zero scatter.

For each primary-secondary pair, we can write the phase-space probability as follows:
\begin{align}\label{eqn:PdVRp_LNz}
    &P(\dVij, \Rij  | \Lci, \zci,  \Nsi )
    \\
    &= \int\rmd M \, P(M|\Lci,\zci,\Nsi) \, P(\dVij, \Rij |M,\Lci,\zci)\,. \nonumber
\end{align}
This represents a marginalization over host halo mass, which serves as a latent variable for each individual primary. 

Using Bayes theorem we can rewrite the probability $P(M|\Lci, \zci,  \Nsi)$, which relates to the galaxy-halo connection, (dropping subscripts for brevity)
\begin{equation}
    P(M|L,z,N) = \dfrac{P(N | M,L,z) \, P(L|M,z)\, n(M,z)}{\int \rmd M \,\, P(N | M,L,z) \, P(L|M,z)\, n(M,z)}\,.
\end{equation}
Here, $n(M,z)$ is the halo mass function at redshift $z$\footnote{As shown in \papII, the expression also involves a completeness factor, but since it drops out of the final expression we have omitted it here for the sake of clarity}. As detailed in \papII, each component in this expression can be computed once the probabilistic link between galaxy luminosity and halo mass, that is, the CLF, is defined. 

The probability $P(\dVij, \Rij |M, \Lci, \zci)$ specifies the phase-space distribution of secondaries in a halo of mass $M$ at redshift $\zci$, with a primary of luminosity $\Lci$. Since the secondary is either a true satellite or an interloper, we write (omitting subscripts for brevity)
\begin{align}
   P(\Delta V, & R_\rmp | M, L,z) = f_{\rm int} (M,L,z) \, P_{\rm int}(\Delta V, R_\rmp | M,L,z) \, + \, 
    \nonumber \\
    &\big[1-f_{\rm int} (M,L,z) \big] \, P_{\rm sat}(\Delta V, R_\rmp | M,L,z)\,.
\end{align}
Here, $f_{\rm int}(M,L,z)$ is the probability for a given secondary to be an interloper given the halo mass and the primary's redshift and luminosity. $P_{\rm int}$ and $P_{\rm sat}$ are the distinct projected phase-space distributions of interlopers and satellites, respectively.

The interloper fraction, $f_{\rm int}(M,L,z)$ is computed as $\lambda_{\rm int}/(\lambda_{\rm int}+\lambda_{\rm sat})$, where the expected mean number of selected satellites, $\lambda_{\rm sat}$, follows from the galaxy-halo connection and forward-modeling of all selection effects. The expected mean number of interlopers is modeled as
\begin{equation} \label{eqn:N_int}
    \lambda_{\rm int}(L,z) = n_{\rm gal}(L,z) \, V_{\rm cone} (L,z) \, b_{\rm eff}(L,z)\,.
\end{equation} 
Here, $n_{\rm gal}$ is the galaxy number density in the luminosity range between $L_{\rm pri}$ and the luminosity threshold of that galaxy group given its redshift, $V_{\rm cone}$ is the volume of the secondary selection cone around that specific primary, and $b_{\rm eff}$ is an effective bias parameter that accounts for the fact that the selection volume around a primary is not a randomly placed volume in the Universe. For the latter, we assume a simple functional form that has three free nuisance parameters that fully specify our interloper bias model and that are constrained simultaneously with all other physical parameters. 

To calculate the projected phase-space density of the true satellites, we use
\begin{align}
P_{\rm sat}(\Delta V, \Rp |& M, L, z) = \\
& P_{\rm sat}(\Delta V | \Rp, M, L, z) \, P_{\rm sat}(\Rp | M, L, z) \nonumber \,.
\end{align}
The probability for the projected separation $P_{\rm sat}(\Rp | M,L,z)$ is fully determined by assuming a radial profile of satellites, $n_{\rm sat}(r|M)$, and by accounting for the primary's redshift and luminosity-dependent selection criteria. For the radial distribution, we use a generalized NFW profile:
\begin{equation}
    n_{\rm sat}(r|M) \propto \bigg( \dfrac{r}{\calR r_\rms} \bigg)^{-\gamma} \, \bigg(1 + \dfrac{r}{\calR r_\rms} \bigg)^{-3 +\gamma}\,,
\end{equation}
where $\calR$ is the ratio of the scale radius of the satellite distribution to that of dark matter, $r_\rms$. Both $\gamma$ and $\calR$ are free parameters that are also fit with \Basilisc.

Given a projected separation, the line-of-sight velocity profile of the satellites, $P_{\rm sat}(\Delta V | \Rp, M,L,z)$, is modeled using a generalized Gaussian distribution, whose variance and kurtosis are computed from second- and fourth-order spherical Jeans equations, as described in \papII. This modeling is extended out to the halo splashback radius to account for in-falling or outgoing back-splash galaxies.

Finally, for the projected phase-space distribution of interlopers beyond the splashback radius, we assume a uniform projected surface density to compute $P_{\rm int}(R_{\rm p}|M,L,z)$ while for the velocity distribution we write $P_{\rm int}(\Delta V |R_{\rm p},M,L,z) \equiv P_{\rm bg} + P_{\rm inf}$. The first component, $P_{\rm  bg}$, describes the distributions of foreground and background galaxies that are kinematically decoupled from the host halo in question, which implies that $P_{\rm bg}(\Delta V)$ is simply proportional to the differential volume of the secondary selection cone corresponding to the velocity slice at $\Delta V$. The other component, $P_{\rm inf}$, describes the population of interlopers that are kinematically coupled to the halo in question due to large-scale infall motion. Their velocity distribution is modeled semi-empirically, guided by the observed line-of-sight velocity distribution of galaxies in the direct neighborhood of, but outside of the halo of, the primary in question. We refer the interested reader to \paperII for more details.

\subsection{Satellite Abundance modeling}
\label{ssec:Ns_model}

The likelihood of the satellite abundances is given by
\begin{equation}
    \calL_{\rm Ns} = P(\bD_{\rm Ns}|\btheta) = \prod\limits_{i=1}^{N_{\rm Ns}} P(\Nsi | \Lci,\zci)\,,
\end{equation}
which makes the reasonable assumption that the halo occupation statistics of different galaxy groups are independent of each other. Each term inside the product over all $N_{\rm Ns}$ primaries can be rewritten as a halo mass marginalization,
\begin{equation} \label{eqn:P_N_given_Lz}
    P(N | L,z) = \int \rmd M \, P(M|L,z) \, P(N | M, L, z)\,.
\end{equation}
Using Bayes' theorem, the first factor in the integrand can be written as
\begin{equation}
    P(M|L,z) = \dfrac{P(L|M,z) \, n(M,z)}{\int \rmd M \, P(L|M,z) \, n(M,z)}\,,
\end{equation}
where $P(L|M,z)$ is fully determined by the CLF. In order to compute $P(N | M, L, z)$ we assume that the occupation of secondaries is a Poisson process, which implies that
\begin{equation}
    P(N|M,L,z) = \dfrac{\lambda_{\rm tot}^{N} \exp(-\lambda_{\rm tot})}{N !}\,.
\end{equation}
Here, $\lambda_{\rm tot}$ is the expectation value for the total number of secondaries (satellites plus interlopers). Using a correction factor, $f_{\rm corr}$, that is specific to each primary galaxy, \Basilisk corrects for aperture and spectroscopic incompleteness by computing $\lambda_{\rm tot} = f_{\rm corr}\,(\lambda_{\rm sat}+\lambda_{\rm int})$. Here, $\lambda_{\rm sat}$ is calculated by integrating $\Phi_\rms(L|M)$, while $\lambda_{\rm int}$ is given by equation~(\ref{eqn:N_int}).

\subsection{Luminosity Function modeling}
\label{ssec:LF_model}

The final observational constraint used by \Basilisk to constrain the galaxy-dark matter connection is the comoving number density of SDSS galaxies, $n_{\rm gal}(L_1,L_2)$ in ten 0.15~dex bins in luminosity, $[L_1,L_2]$  covering the range $9.5 \leq \log L/(h^{-2} \Lsun) \leq 11.0$ (see \S\ref{sec:data}). The log-likelihood for these data, to which we loosely refer as the LF-data, is computed using
\begin{equation}
\ln\calL_{\rm LF}(\bn_{\rm obs}|\btheta) = -\frac{1}{2} \, [\bn(\btheta) - \bn_{\rm obs}]^t \, \bC^{-1}_{ij}  \, [\bn(\btheta) - \bn_{\rm obs}]\,.
\end{equation}
Here, $\bn_{\rm obs}$ is the data vector and $\bn(\btheta)$ is the corresponding model prediction, calculated from the CLF using
\begin{equation}\label{numdens}
n_{\rm gal}(L_1,L_2) = \int_{L_1}^{L_2} \rmd L \int_{0}^{\infty} \Phi(L|M) \, n(M,z_{\rm SDSS}) \, \rmd M\,,
\end{equation}
with $z_{\rm SDSS} = 0.1$ the characteristic redshift for the SDSS data used and $\bC$ the covariance matrix of the data (see \paperII for details).

\section{The Galaxy-Halo Connection Models}
\label{sec:GHC_models}

Here we describe the various CLF models that we use to characterize the galaxy-halo connection. We start by first describing the generic form of the CLF model used in this work, which has been the standard form for almost two decades (see Section~\ref{sec:intro}). Next, we define different model variations, (A) to (F), which we implement in \Basilisc. These are listed in the order of increasing flexibility (increasing degrees of freedom). Note that the sequential opening of model freedom is cumulative, that is, for example, the model freedom allowed in Model~(D) is in addition to all model flexibilities introduced in Model~(C), etc.

\subsection{Conditional Luminosity Function}
\label{ssec:CLF}

The conditional luminosity function (CLF), $\Phi(L|M,z)$, specifies the average number of galaxies with luminosities in the range $[L -\rmd L, L+\rmd L]$ that reside in a halo of viral mass $M$ at redshift $z$. As is common, we split the CLF in its central and satellite components, i.e.,
\begin{equation}
    \Phi(L|M,z) = \Phi_\rmc(L|M) + \Phi_\rms(L|M)\,,
\end{equation}
where the subscripts ‘c’ and ‘s’ refer to central and satellite, respectively. We assume that the luminosities of galaxies in any given halo are independent draws from the CLF, and that the CLF is independent of the redshift over the range considered here ($0.02 \leq z \leq 0.2$).

The probability distribution of central galaxy luminosity is assumed to be a log-normal distribution,
\begin{equation}\label{eqn:CLFcen}
\Phi_\rmc (L | M) \rmd L = \frac{\log e}{\sqrt{2\pi \sigma_\rmc^2}} \exp \left[ -\left(\frac{\log L - \log\bar{L}_\rmc}{\sqrt{2} \sigma_\rmc} \right)^2\right] \frac{\rmd L}{L}\,.
\end{equation}
This functional form is motivated by what is observed in galaxy group catalogs \citep[][]{Yang.etal.09} and predicted by semi-analytical models of galaxy formation \citep[][]{Zheng.etal.05}. The median luminosity, $\bar{L}_\rmc$, is parametrized to vary as a double power law of the host halo mass, in agreement with the consensus SHMR of central galaxies \citep[see][]{Wechsler.Tinker.18}. Hence,
\begin{equation}\label{eqn:averLc}
\bar{L}_\rmc (M) = L_0 \frac{(M / M_1)^{\gamma_1}}{(1 + M / M_1)^{\gamma_1 - \gamma_2}}\,.
\end{equation}
This relation between halo mass and median central galaxy luminosity is determined by four free parameters: a normalization, $L_0$, a characteristic halo mass, $M_1$, and two power-law slopes, $\gamma_1$ and $\gamma_2$. The parameter $\sigma_\rmc$ controls the width of the log-normal distribution, that is, the scatter in the luminosity of central galaxies for halos of a given mass.

For the satellite CLF we adopt a modified Schechter function, motivated by structure formation theory \citep{Press.Schechter.1974} and findings in galaxy group catalogs \citep{Yang.etal.09}: 
\begin{equation}\label{eqn:satCLF}
\Phi_\rms (L | M) = \frac{\phi_\rms^*}{L_\rms^*} \left( \frac{L}{L_\rms^*} \right)^{\alpha_\rms} \exp \left[ - \left( \frac{L}{L_\rms^*} \right)^2 \right]\,.
\end{equation}
Thus, the luminosity distribution of satellites, in a halo of a given mass, is assumed to be a power law with slope $\alpha_\rms$ and with an exponential cut-off above a critical luminosity, $L_\rms^*$. It has become standard to model the mass dependence of the satellite CLF normalization as 
\begin{equation}\label{eqn:satCLFnorm}
\log \left[ \phi_\rms^*(M) \right] = b_0 + b_1 \log M_{12} + b_2 (\log M_{12})^2\,,
\end{equation}
where $M_{12} = M/(10^{12}\Msunh)$, which is the parametrization we adopt throughout. In what follows, we will define $L_\rms^*$ in terms of the corresponding median central galaxy luminosity, $\bar{L}_\rmc$, through 
\begin{equation}\label{eqn:deltas}
\Delta_\rms(M) \equiv \log L_\rms^*(M) - \log \bar{L}_\rmc(M)\,.
\end{equation}
The different models for the CLF discussed below all follow this specific form for the CLF, but differ in the way that the various parameters, in particular $\sigma_\rmc$, $\alpha_\rms$, and $\Delta_\rms$ scale with halo mass.
\begin{figure*}
\centering
\includegraphics[width=0.9\textwidth]{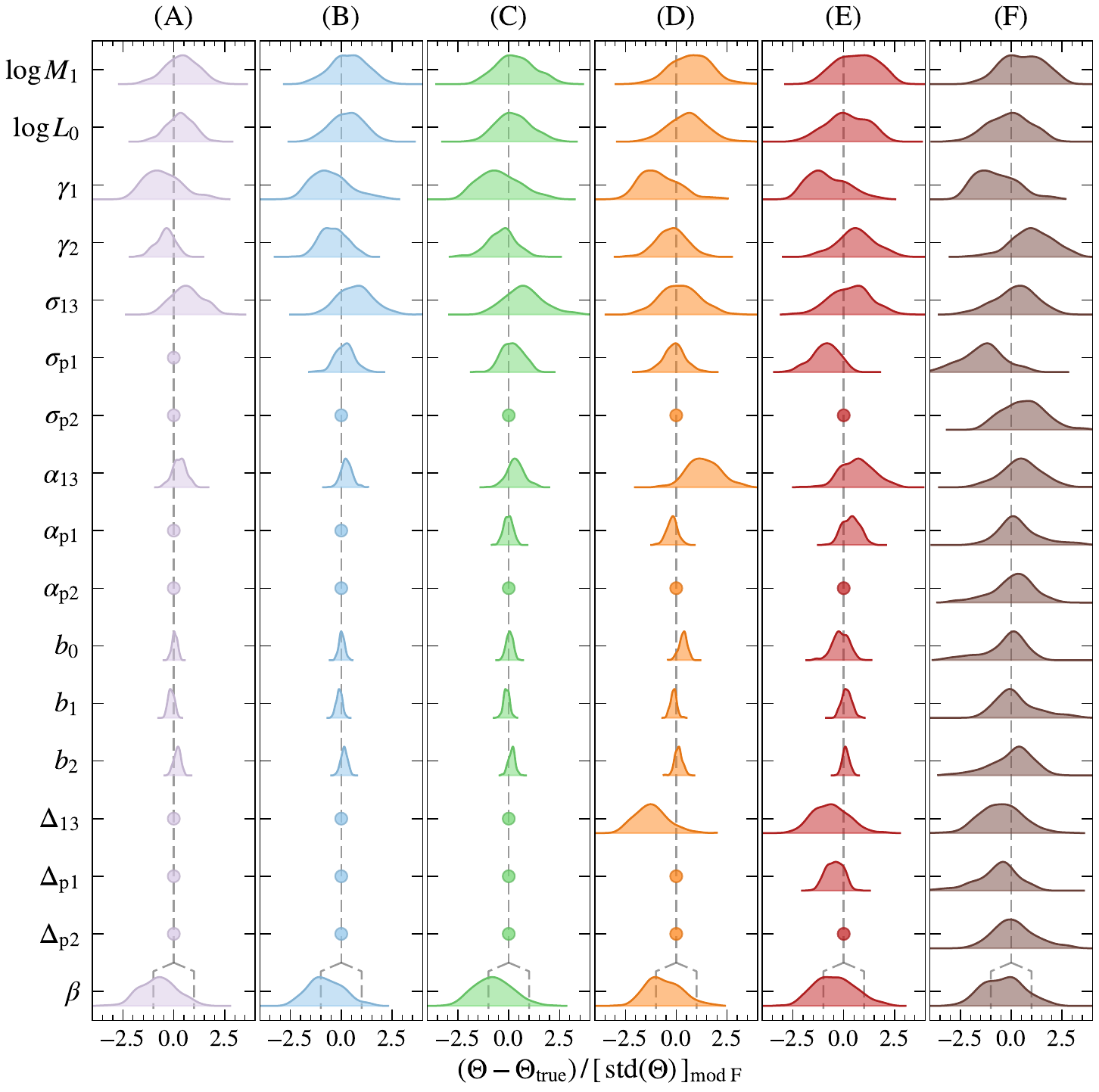}
\caption{Posterior distributions of the various parameters for the galaxy-halo connection obtained by fitting the mock survey data. Different columns correspond to different CLF models used, as indicated at the top. Note that each parameters is indicated with respect to its true value that was used to create the mock data (indicated by the vertical, gray-dashed line), and scaled by the $1 \sigma$ width of the posterior of model (F). The latter facilitates a meaningful comparison of the widths of the posterior distributions among different CLF models. Whenever a parameter is kept fixed at its true value, this is indicated by a circle; note how going from model (A) to (F), fewer and fewer model parameters are kept fixed, which causes most posterior distributions to widen. Finally, the parameter, $\beta$, shown at the bottom, represents the mean orbital velocity anisotropy of satellite galaxies. Here the gray line splits into a range that marks the 16-84 percentile range of velocity anisotropies of subhalos in individual host halos in the simulation box that was used to create the mock data. Note how, in this validation test, all parameters show unbiased recovery, irrespective of the different CLF models assumed.}
\label{fig:par_recovery}
\end{figure*}

\subsubsection{Model (A): Basic CLF}
\label{sec:modelA}

For our simplest and most restrictive CLF model, we assume that $\sigma_\rmc$, $\alpha_\rms$, and $\Delta_\rms$ are all independent of halo mass. While $\sigma_\rmc$ and $\alpha_\rms$ are treated as free parameters, we keep $\Delta_\rms$ fixed to a value of $-0.25$.  The latter implies that $L_\rms^*(M) = 0.562 \bar{L}_\rmc(M)$, which is motivated by the results from galaxy group catalogs \citep[][]{Yang.etal.09}. As a result, $\Phi_\rmc(L|M)$ and $\Phi_\rms(L|M)$ are characterized by 5 and 4 free parameters, respectively, totaling 9 parameters for the entire CLF. This is a fairly standard CLF model that was used, for example, in \citet{Cacciato.etal.13}.  It is also the CLF parametrization that we use to construct mock SDSS data for testing and validation in Papers~I and~II (see also Section~\ref{sec:mock_validation}).

\subsubsection{Model (B): Mass-dependent scatter in central luminosity}
\label{sec:modelB}

In our next model, we introduce additional freedom by allowing the scatter in the luminosities of central galaxies to depend linearly on halo mass, i.e., we set
\begin{equation}
\sigma_\rmc(M) = \sigma_{13} + \sigma_{\rm p1}\log M_{13}\,,
\end{equation}
where $M_{13}$ is the virial mass of the halo in units of $10^{13} \Msunh$. Hence, compared to model (A), we add one additional parameter, $\sigma_{\rm p1}$. Since there is no a priori reason for the central luminosity scatter to be independent of mass \citep[][]{Wechsler.Tinker.18} this is a reasonable first-order modification. Note that this CLF model has been adopted in several previous studies \citep[][]{Lange.etal.19a, Lange.etal.19b, vdBosch.etal.19}, including our own \Basilisc-based analysis of the SDSS data presented in \papII. To protect against unphysical models, we set $\sigma_\rmc(M)=0$ for any halo mass for which the above equation implies $\sigma_\rmc < 0$.

\subsubsection{Model (C): Mass-dependent satellite faint-end slope}
\label{sec:modelC}

Next, we also allow for a mass dependence of the faint-end slope of the satellite CLF, i.e., we set
\begin{equation} \label{loglinalpha}
\alpha_\rms(M) = \alpha_{13} + \alpha_{\rm p1} \log M_{13}\,,
\end{equation}
which introduces one additional parameter, $\alpha_{\rm p1}$, characterizing the log-linear slope of this mass dependence. Interestingly, several of the early CLF-based analyses allowed for a similar mass dependence of $\alpha_\rms$ \citep[][]{Yang.etal.03, vdBosch.etal.07}. These studies always inferred $\alpha_{\rm p1}$ to be negative, implying that the faint-end slope of the satellite CLF becomes steeper with increasing halo mass. However, this freedom in the CLF model was abandoned by \citet{Cacciato.etal.13} in order to make their MCMC chains more stable and converge better. Most subsequent CLF-based studies followed suit and assumed $\alpha_\rms$ to be halo mass independent. To guard against CLFs for which the total integrated luminosity is unbounded, we set $\alpha_\rms = -2$ for any halo mass for which the above equation implies $\alpha_\rms < -2$.

\subsubsection{Model (D): Flexible high-luminosity cutoff for satellites}
\label{sec:modelD}

In this model, we build on model (C) by allowing $\Delta_\rms$ to be a free parameter, that is, we no longer keep it fixed at $-0.25$. The latter value was motivated by direct fits to the CLF inferred from galaxy group catalogs by \citet{Yang.etal.08}. However, the main reason why these authors adopted a fixed mass-independent value was to reduce the number of free parameters to more reliably fit the noisy data. Note also that the CLF inferred from the group catalog is based on the assumption that the central galaxy is the most luminous galaxy in the group, which is not always the case \citep[][]{vdBosch.etal.05b, Skibba.etal.11, Hoshino.etal.15, Lange.etal.18}. \Basilisk accounts for this through impurity modeling, by accounting for the probability that a satellite outshines its corresponding central galaxy (see \paperII for details). As discussed in more detail in Section~\ref{sec:discussion}, whether satellites are allowed to be the brightest halo galaxies (BHG) or not has important implications for the inferred galaxy-halo connection.

\subsubsection{Model (E): Mass-dependent satellite cut-off luminosity}
\label{sec:modelE}

This model is an obvious extension of model (D), in which we introduce a mass dependence in $\Delta_\rms$. In particular, we define
\begin{equation}
\Delta_\rms(M) = \Delta_{13} + \Delta_{\rm p1} \log M_{13}\,,
\end{equation}
which introduces one additional parameter, $\Delta_{\rm p1}$, which specifies the log-linear slope.

\subsubsection{Model (F): Quadratic mass dependencies}
\label{sec:modelF}

For our final, most flexible model, we assume that the mass dependencies of $\sigma_\rmc$, $\alpha_\rms$, and $\Delta_\rms$ are all quadratic, rather than linear. In particular, we introduce three new parameters, $\sigma_{\rm p2}$, $\alpha_{\rm p2}$, and $\Delta_{\rm p2}$ that characterize the quadratic behavior according to
\begin{eqnarray}
\sigma_\rms(M) & = \sigma_{13} + \sigma_{\rm p1} \log M_{13} + \sigma_{\rm p2} (\log M_{13})^2\,,\nonumber \\
\alpha_\rms(M) & = \alpha_{13} + \alpha_{\rm p1} \log M_{13} + \alpha_{\rm p2} (\log M_{13})^2\,,\nonumber \\
\Delta_\rms(M) & = \Delta_{13} + \Delta_{\rm p1} \log M_{13} + \Delta_{\rm p2} (\log M_{13})^2\,.
\end{eqnarray}
As a result, in this most flexible model, $\Phi_\rmc$ and $\Phi_\rms$ are characterized by a total of 16 free parameters; 7 for $\Phi_\rmc(L|M)$ and 9 for $\Phi_\rms(L|M)$. This is 7 parameters more than the basic CLF of model (A).
\begin{figure*}
\centering
\includegraphics[width=0.9\textwidth]{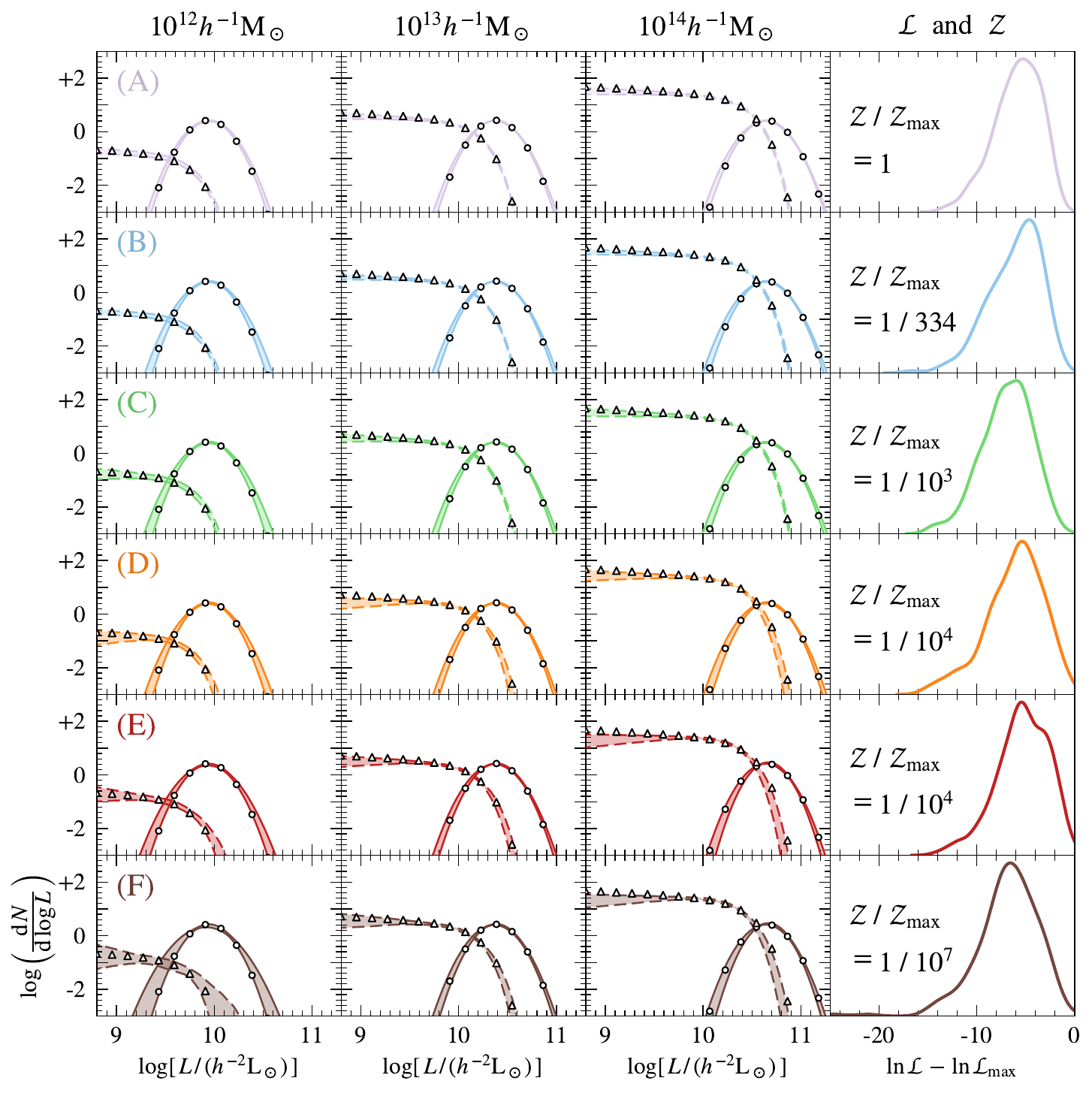}
\caption{The conditional luminosity function of galaxies in the mock survey data, their relative likelihood distributions, and the Bayes factor corresponding to different CLF model assumptions. The first 3 columns show the CLFs for different halo masses, as indicated at the top. Different rows show the results for the six different CLF models used by \Basilisk for the inference, as indicate in the left-most panels. The symbols (circles/triangles) show the true input CLF (of centrals/satellites) used to create the mock, while the shaded regions indicate the $1\sigma$ confidence intervals based on \Basilisc's inference. The histograms in the rightmost column show the corresponding total log likelihood distributions of the posteriors, relative to the maximum likelihood estimate among all models tested. The values of $\calZ / \calZ_{\rm max}$ that are indicated in the rightmost panels indicate the Bayes Factor for the inference of each CLF model with respect to the model with highest evidence, which in this case is model (A).}
\label{fig:CLF_mock}
\end{figure*}

\subsection{Bayes Factors}
\label{sec:bayes}

In order to meaningfully compare the likelihoods obtained using different models, we rely on the Bayes factor. This properly penalizes an increase in model freedom that does not result in a significant improvement in the likelihood. The Bayes factor for two models 1 and 2 is defined as the ratio of their evidences, $\calZ_2/\calZ_1$, where the Bayesian evidence is given by
\begin{equation}\label{eqn:evidence_defn}
\mathcal{Z} = \int \rmd \btheta \, P(\btheta) \, \calL(\bD|\btheta) \,.
\end{equation}
Here $P(\btheta)$ is the prior on the set of all model parameters, $\btheta$, and $\calL(\bD|\btheta)$ is the likelihood of the data $\bD$ given the model. For each model, we compute the evidence using the method outlined in Appendix~\ref{App:evidence_calculation}, using the likelihood distribution obtained with \Basilisc. In general, models for which the evidence is larger are preferred. Hence, if the Bayes factor $\calZ_1/\calZ_2$ is smaller than unity, then model 2 is to be preferred over 1, and vice versa. 

\section{Application to Mock Data}
\label{sec:mock_validation}

For each of the six CLF models described above, we apply \Basilisk to the mock data described in Section~\ref{sec:mockdata}. Recall that the mock data are constructed assuming a CLF parametrization that corresponds to our most restrictive model~(A). Hence, the mock data are based on a CLF for which $\sigma_{\rm p1} = \sigma_{\rm p2} = \alpha_{\rm p1} = \alpha_{\rm p2} = \Delta_{\rm p1} = \Delta_{\rm p2} = 0$ and $\Delta_\rms = -0.25$. 
\renewcommand{\arraystretch}{2}
\begin{table*}[t!]\label{tab:param}
\centering
\begin{tabular}{c| r@{.}l r@{.}l r@{.}l r@{.}l r@{.}l r@{.}l}
 \hline
 Parameters & \multicolumn{2}{c}{Model A} & \multicolumn{2}{c}{Model B} & \multicolumn{2}{c}{Model C} & \multicolumn{2}{c}{Model D} & \multicolumn{2}{c}{Model E} & \multicolumn{2}{c}{Model F} \\ 
 \hline\hline
 $\log M_1$ & 11&$37^{+0.08}_{-0.13}$ & 11&$40^{+0.11}_{-0.11}$ & 11&$17^{+0.14}_{-0.14}$ & 10&$94^{+0.05}_{-0.04}$ & 10&$99^{+0.08}_{-0.04}$ & 10&$95^{+0.06}_{-0.06}$\\ 
 $\log L_0$ & 10&$03^{+0.03}_{-0.04}$ & 10&$04^{0.04}_{-0.03}$ & 9&$94^{+0.05}_{-0.05}$ & 9&$89^{+0.03}_{-0.02}$ & 9&$94^{+0.04}_{-0.03}$ & 9&$83^{+0.05}_{-0.04}$\\
 $\gamma_1$ & 2&$38^{+0.48}_{-0.25}$ & 2&$30^{+0.44}_{-0.31}$ & 3&$17^{+0.79}_{-0.61}$ & 4&$63^{+0.25}_{-0.40}$ & 4&$44^{+0.39}_{-0.52}$ & 4&$41^{+0.39}_{-0.38}$\\
 $\gamma_2$ & 0&$21^{+0.006}_{-0.006}$ & 0&$21^{+0.009}_{-0.008}$ & 0&$24^{+0.008}_{-0.008}$ & 0&$20^{+0.009}_{-0.009}$ & 0&$17^{+0.014}_{-0.014}$ & 0&$22^{+0.019}_{-0.017}$\\
 $\sigma_{13}$ & 0&$18^{+0.002}_{-0.002}$ & 0&$18^{+0.002}_{-0.003}$ & 0&$17^{+0.003}_{-0.002}$ & 0&$21^{+0.003}_{-0.003}$ & 0&$21^{+0.004}_{-0.004}$ & 0&$21^{+0.004}_{-0.005}$\\
 $\sigma_{\rm p1}$ & 0&$00^*$ & 0&$002^{+0.003}_{-0.003}$ & $-0$&$010^{+0.003}_{-0.004}$ & $-0$&$008^{+0.005}_{-0.005}$ & $-0$&$008^{+0.005}_{-0.005}$ & $-0$&$015^{+0.005}_{-0.006}$\\ 
 $\sigma_{\rm p2}$ & 0&$00^*$ & 0&$00^*$ & 0&$00^*$ & 0&$00^*$ & 0&$00^*$ & 0&$014^{+0.006}_{-0.007}$\\ 
 $\alpha_{13}$ & $-0$&$82^{+0.05}_{-0.05}$ & $-0$&$82^{+0.05}_{-0.05}$ & $-0$&$54^{+0.07}_{-0.06}$ & $-1$&$59^{+0.05}_{-0.05}$ & $-1$&$55^{+0.06}_{-0.04}$ & $-1$&$58^{+0.06}_{-0.06}$\\
 $\alpha_{\rm p1}$ & 0&$00^*$ & 0&$00^*$ & $-0$&$68^{+0.06}_{-0.08}$ & $-0$&$40^{+0.06}_{-0.06}$ & $-0$&$45^{+0.05}_{-0.08}$ & $-0$&$33^{+0.06}_{-0.07}$\\ 
 $\alpha_{\rm p2}$ & $0$&$00^*$ & $0$&$00^*$ & $0$&$00^*$ & $0$&$00^*$ & $0$&$00^*$ & $-0$&$09^{+0.14}_{-0.08}$\\ 
 $b_0$ & $-0$&$64^{+0.03}_{0.03}$ & $-0$&$65^{+0.03}_{0.03}$ & $-0$&$53^{+0.03}_{-0.03}$ & $-1$&$107^{+0.06}_{-0.06}$ & $-0$&$99^{+0.07}_{-0.06}$ & $-1$&$26^{+0.08}_{-0.10}$\\
 $b_1$ & $1$&$03^{+0.04}_{-0.04}$ & $1$&$03^{+0.03}_{-0.04}$ & $0$&$98^{+0.04}_{-0.04}$ & $0$&$71^{0.06}_{-0.05}$ & $0$&$64^{+0.05}_{-0.05}$ & $0$&$96^{+0.14}_{-0.12}$\\ 
 $b_2$ & $-0$&$10^{+0.01}_{-0.01}$ & $-0$&$10^{+0.01}_{-0.01}$ & $-0$&$14^{+0.02}_{-0.01}$ & $-0$&$04^{+0.02}_{-0.02}$ & $-0$&$04^{+0.01}_{-0.02}$ & $-0$&$12^{+0.05}_{-0.04}$\\
 $\Delta_{13}$ & $-0$&$25^*$ & $-0$&$25^*$ & $-0$&$25^*$ & $0$&$18^{+0.01}_{-0.02}$ & $0$&$18^{+0.01}_{-0.02}$ & $0$&$18^{+0.01}_{-0.02}$\\
 $\Delta_{\rm p1}$ & $0$&$00^*$ & $0$&$00^*$ & $0$&$00^*$ & $0$&$00^*$ & $0$&$08^{+0.03}_{-0.03}$ & $-0$&$09^{+0.04}_{-0.06}$\\ 
 $\Delta_{\rm p2}$ & $0$&$00^*$ & $0$&$00^*$ & $0$&$00^*$ & $0$&$00^*$ & $0$&$00^*$ & $0$&$06^{+0.03}_{-0.03}$\\
 $\beta$ & $0$&$18^{+0.06}_{-0.06}$ & $0$&$18^{+0.06}_{-0.06}$ & $0$&$18^{+0.06}_{-0.06}$ & $0$&$18^{+0.06}_{-0.06}$ & $0$&$18^{+0.06}_{-0.06}$ & $0$&$18^{+0.06}_{-0.06}$\\
 [1ex] 
 \hline
\end{tabular}
\caption{Galaxy-halo connection parameters inferred with \Basilisk using SDSS data for each of the 6 different CLF models (different columns). The quoted values are the medians and the 68\% confidence intervals for the corresponding posterior distributions. Asterisks ($^*$) indicate that that parameter was held fixed to the value indicated.}
\end{table*}

Fig.~\ref{fig:par_recovery} shows the resulting posteriors on the various CLF model parameters, as well as the posterior on the inferred velocity anisotropy parameter, $\beta$. The different columns correspond to the different CLF models used in the analysis, as indicated at the top. The parameter posteriors are shown relative to their true input values used to create the mock survey and are rescaled by the respective standard deviations of the posteriors obtained while fitting the most flexible CLF model (F). This is done for ease of comparison and nicely shows how the posteriors widen as a more flexible model is assumed. The vertical gray dashed line in each panel indicates the true parameter values used to create the mock in these scaled units. However, in the bottom panels, which show the posteriors for $\beta$, the gray dashed lines split to show a range, rather than a singular value. Here, the two gray-dashed lines indicate the $1\sigma$ range of anisotropy values for the subhalos in the simulation used to create the mock. 

In each column, the parameters that are not part of that specific parametrization are shown as circles and are kept fixed at their true values while fitting the mock data. For example, in the first column, which corresponds to model (A), 7 parameters ($\sigma_{\rm p1}$, $\sigma_{\rm p2}$, $\alpha_{\rm p1}$, $\alpha_{\rm p2}$, $\Delta_{13}$, $\Delta_{\rm p1}$, and $\Delta_{\rm p2}$) are kept fixed to their true values ($0$, $0$, $0$, $0$, $-0.25$, $0$, and $0$, respectively) as shown by the circles. Note that whenever $\sigma_\rms$, $\alpha_\rms$ or $\Delta_\rms$ are mass independent, $\sigma_{13}$, $\alpha_{13}$, and $\Delta_{13}$ are used to indicate their values. 
%In the second column, for model (B), $\sigma_{\rm p1}$ is no longer fixed but instead has a posterior distribution, which evidently is in perfect agreement with the true value. 
%
\begin{figure*}
\centering
\includegraphics[width=0.9\textwidth]{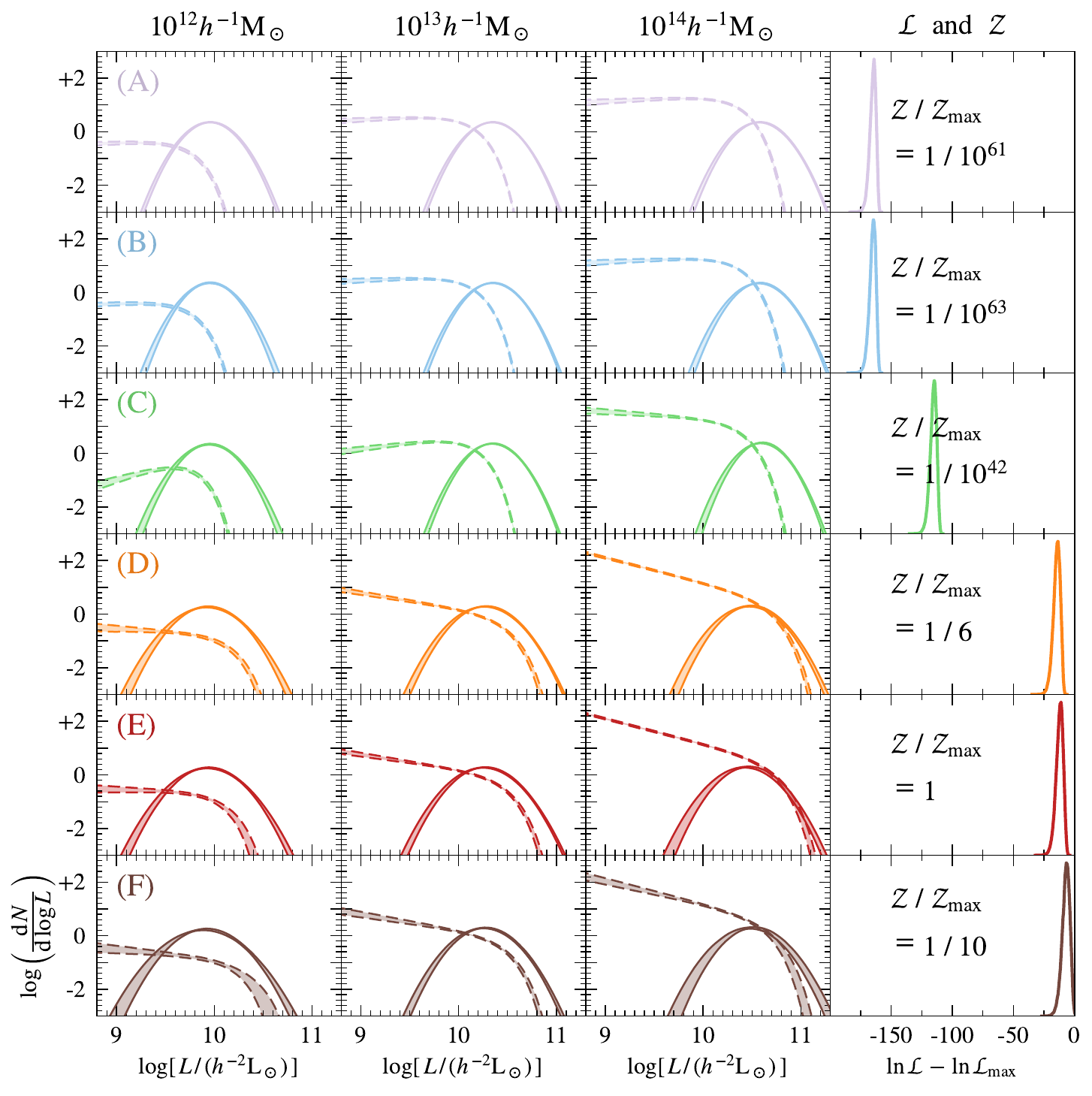}
\caption{Same as Fig.~\ref{fig:CLF_mock}, but for an analysis of the SDSS DR7 data. 
Unlike for the mock data, here the inferred CLFs depend strongly on the CLF model used. In particular, the posteriors of the log likelihood distributions relative to the maximum likelihood estimate among all models, shown in the rightmost panels, shows two dramatic jumps going from model (B) to (C), and from model (C) to (D), indicating that a mass-dependence for $\alpha_\rms$ and freedom in $\Delta_\rms$ are strongly favoured by the data. The Bayes factor with respect to the model with the highest evidence, indicated in the right-most panel, suggests that model (E) is the optimal model to characterize the halo occupation statistics of SDSS galaxies.}
\label{fig:CLF_sdss}
\end{figure*}

If we look at the results for model (A), we see that the posteriors for all 10 parameters are perfectly consistent with the true input values. This validates \Basilisk as a reliable tool for obtaining unbiased constraints on the galaxy-halo connection using the kinematics and abundances of satellite galaxies, combined with the total luminosity function (see also \papII). Moving up to more and more flexible models, we see that the inferred posteriors in each case remain perfectly consistent with the input values (i.e., the posterior distributions never lie too far from the gray-dashed lines). However, with increasing freedom, the posterior distributions for most parameters start to broaden. In particular, for the most flexible model (F) shown in the right-most column, the posterior distributions for several parameters (notably $\alpha_{13}$, $b_0$, $b_1$, and $b_2$) are significantly broader than in the case of model (A). This is exactly the behavior to be expected if \Basilisk yields unbiased inference: when using a model identical to that used to create the mock data, here model (A), then all parameters are perfectly recovered, and when allowing for more freedom in the model than required, the posteriors broaden due to covariance among the model parameters, but the inference remains unbiased. Importantly, the additional freedom, which goes beyond what is inherent in the data, does not cause any systematic error in the inference.

Fig.~\ref{fig:CLF_mock} summarizes the key results from the mock-based test. This time, the six different models are shown as different rows. The first three columns show the conditional luminosity distribution of central and satellite galaxies in terms of ${\rm d}N_{\rm gal}/ {\rm d}\log L$, for three different halo masses ($10^{12}$, $10^{13}$, and $10^{14} \Msunh$) as indicated at the top. The circles and triangles indicate the true CLFs of centrals and satellites, respectively, that is, the CLFs used to populate the halos while creating the mock galaxy survey. The colored shaded regions indicate the $1\sigma$ confidence intervals of the CLFs predicted based on \Basilisc's inference. Note the perfect recovery of the input CLFs, irrespective of the choice of the model. Note also how with increasing model freedom, the $1\sigma$ confidence intervals broaden as a consequence of the broadening of the parameter posteriors. The broadening is less prominent for $10^{13} \Msunh$ as that is the mass range where the constraints are tightest. 

The fourth column of Fig.~\ref{fig:CLF_mock} shows the posterior distributions of the total likelihoods ($\calL_{\rm SK}+\calL_{\rm Ns}+\calL_{\rm LF}$) of each model with respect to the maximum likelihood estimate among all models tested here. Note that all these distributions are virtually overlapping; the likelihoods do not improve with increasing model flexibility. In each panel, we also indicate the value of the Bayes factor, $\calZ/\calZ_{\rm max}$, where $\calZ_{\rm max}$ is the maximum evidence among all six models. This ratio is unity for model (A), indicating that model (A) has the largest evidence. This is reassuring, since this was the model used to construct the mock data. Note how the increased freedom of models (B) to (F) causes an ever decreasing Bayes factor; the most flexible model (F) has an evidence that is 7 orders of magnitude smaller than that of model (A); clearly models (B) to (F) are severely penalized for being overly flexible. All of this implies that \Basilisc's inference is robust and unbiased, and that the excess model freedom does not artificially improve the quality of the fit to the data. Clearly, \Basilisk not only has the ability to break parameter degeneracies and fit the data with overly flexible models, but also has the power to test where any model extension is justified and required by the data.
\begin{figure*}
\centering
\includegraphics[width=0.85\textwidth]{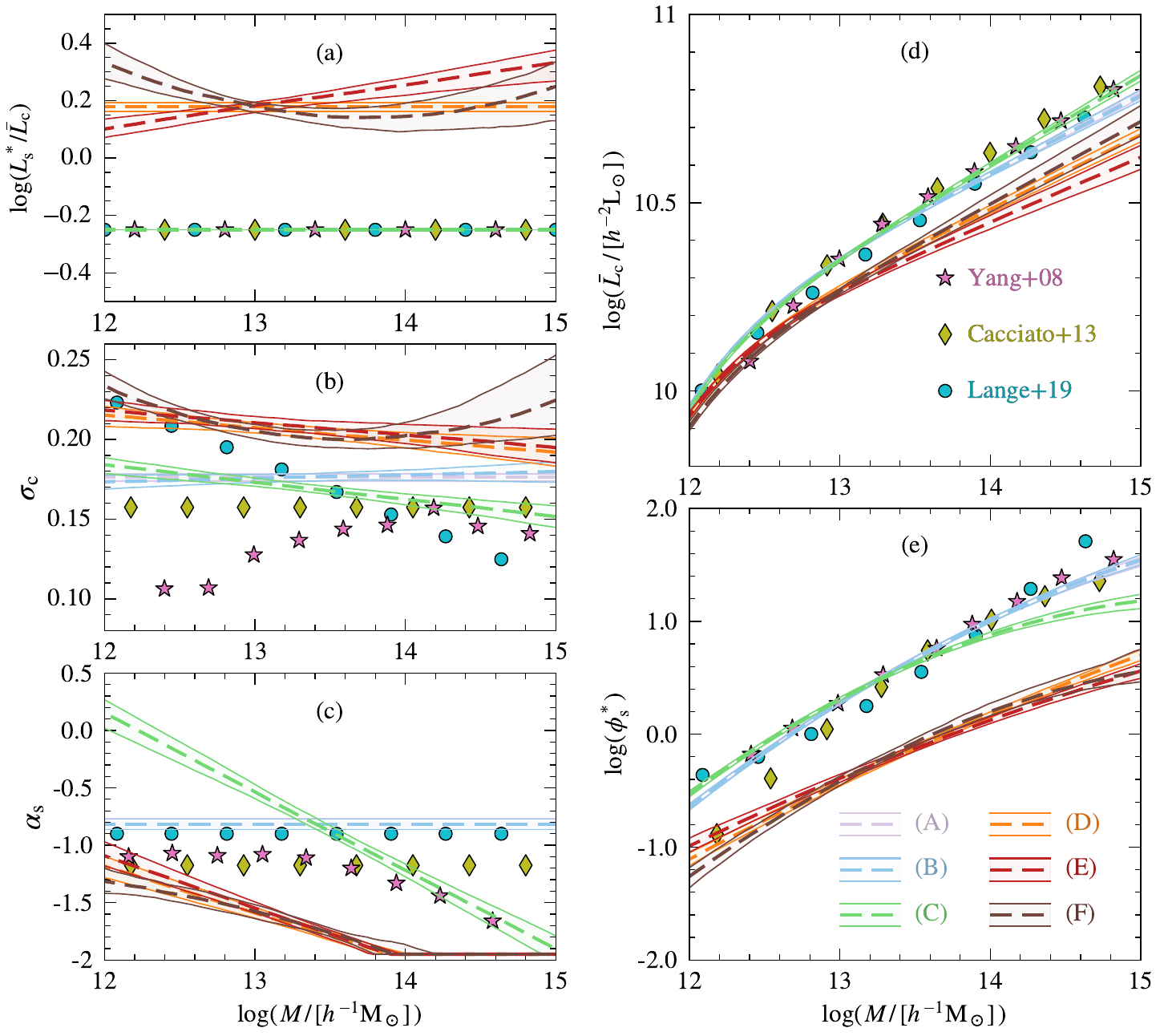}
\caption{Halo mass dependence of different CLF properties. Dashed lines and shaded regions of corresponding color indicate the posterior constraints (68 percentile ranges) obtained with \Basilisk analyzing the SDSS data for the six different CLF models (A) to (F), as indicated in the bottom-right panel. For comparison, the different symbols indicate the results obtained previously using galaxy group catalogs  \citep{Yang.etal.08}, a combination of galaxy clustering and weak lensing \citep{Cacciato.etal.13}, and an independent analysis of satellite kinematics \citep{Lange.etal.19b}, as indicated in the top-right panel. \textit{Panel (a)}: The logarithm of the ratio of the characteristic satellite luminosity and the median luminosity of centrals. \textit{Panel (b)}: The intrinsic scatter in central galaxy luminosity. \textit{Panel (c)}: The faint-end slope of the satellite CLF. \textit{Panel (d)}: The median central galaxy luminosity. \textit{Panel (e)}: The normalization of the satellite CLF. See text for a detail discussion.}
\label{fig:GHC_pars_sdss}
\end{figure*}

\section{Results From The SDSS Data}
\label{sec:result}

We now apply \Basilisk to the SDSS DR7 data described in Section~\ref{sec:data}, using the same set of six CLF models.  
The resulting posterior constraints for all CLF parameters, as well as the orbital anisotropy parameter $\beta$, are listed in Table~1. Fig.~\ref{fig:CLF_sdss} summarizes the key results, similar to Fig.~\ref{fig:CLF_mock} for the mock data. The different rows correspond to different CLF models, as indicated. The first three columns show the CLFs for centrals (solid lines) and satellites (dashed lines) for three different halo masses, indicated at the top, with the shaded colored regions marking the $1\sigma$ confidence intervals obtained with \Basilisc. The colored curves in the fourth column indicate the distributions of $\ln \calL - \ln\calL_{\rm max}$, where $\calL_{\rm max}$ is the maximum likelihood among all six models. Each panel also indicates the Bayes factor, $\calZ/\calZ_{\rm max}$, where as before $\calZ_{\rm max}$ is the maximum evidence among all six models. 

Going from model~(A) to model~(B), which reflects adding a mass dependence to the scatter of the central galaxy luminosities, the CLFs that are inferred are virtually indistinguishable, and the likelihood does not improve at all. In fact, the slope is inferred to be $\sigma_{\rm p1} = 0.002 \pm 0.003$ (see Table~1), which is consistent with zero (see also \papII). Consequently, the Bayes factor $\calZ_{(B)}/\calZ_{(A)}$ is $\sim 10^{-2}$, indicating that model~(A) is to be preferred over model~(B). 

Next, going from model (B) to model (C), which includes a mass dependence for the faint-end slope of the satellite CLF, the evidence increases drastically, with a Bayes factor $\calZ_{(C)}/\calZ_{(B)} \simeq 10^{21}$; apparently the SDSS data strongly prefer a mass-dependent faint-end slope with $\alpha_{\rm p1} \simeq -0.68$.
%, which implies that $\alpha_\rms$ increases from $-1.9$ for massive clusters to $\alpha_\rms \geq 0$ for halos with $M \lta 10^{12.2} \Msunh$. 
This mass dependence is clearly evident from the slopes of the satellite CLFs shown in the first 3 panels in row (C) of Fig.~\ref{fig:CLF_sdss}. 
\begin{figure}
    \centering
    \includegraphics[width=0.95\linewidth]{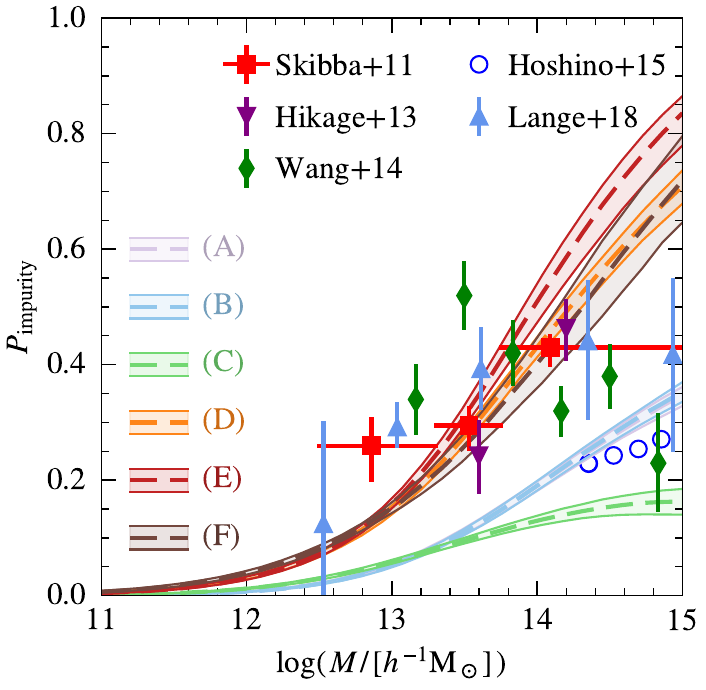}
    \caption{The probability that the brightest halo galaxy (BHG) in a halo is not the central galaxy as a function of halo mass. The different dashed lines with shaded regions show the median relations and their $1\sigma$ confidence intervals, based on \Basilisc's inference of the SDSS data using the six different CLF models, as indicated. Different symbols indicate various observational constraints, as indicated.}
    \label{fig:cen_bhg}
\end{figure}

Treating $\Delta_\rms$ as an additional free parameter in model (D) gives rise to another drastic increase in evidence, and thus a significant improvement in the quality of fit to the data, with $\calZ_{(D)}/\calZ_{(C)} \sim 10^{41}$. Rather than the value of $-0.25$ adopted in models (A) - (C), which was motivated by the results of \citet{Yang.etal.08}, the best-fit value for $\Delta_\rms$ in model (D) is $+0.18$ (see Table~1). Freedom in the $(L_\rms^*/L_\rmc)$ ratio has previously been unexplored, and we find that the SDSS data strongly prefer it. The unbiased recovery of the CLF in the mock test with this model flexibility gives us confidence that this departure of $\Delta_\rms$ from the fiducial value of $-0.25$, is real and is strongly demanded by the data. Interestingly, the faint-end slopes of the satellite CLFs are inferred to be significantly steeper than for model~(C). See 
Section~\ref{sec:implications} below for a more detailed discussion.
%with $\alpha_\rms$ decreasing from $-1.2$ for Milky-Way size halos to $\alpha_\rms = -2$ for $M \gta 10^{14} \Msunh$ (recall from Section~\ref{sec:modelC} that we require $\alpha_\rms \geq -2$ to guard against models for which the total integrated satellite luminosity is infinite). 

In model (E), allowing for mass dependence in $\Delta_\rms$, only a very modest improvement in the evidence is achieved. With $\calZ_{(E)}/\calZ_{(D)} = 6$, model (E) is only weakly preferred over model (D), and the actual posterior predictions of the CLFs in model (E) are difficult to distinguish from those in model (D). The best-fit value for the inferred slope of the mass dependence of $\Delta_\rms$ is $\Delta_{\rm p1} = 0.08 \pm 0.03$ (see Table~1), which is marginally consistent with zero. Finally, going from model (E) to (F), which changes the halo mass dependencies of $\sigma_\rmc$, $\alpha_\rms$ and $\Delta_\rms$ from linear to quadratic by adding three new parameters, is not warranted by the data, as the Bayesian evidence decreases by a factor of ten.  Indeed, all three parameters characterizing the quadratic mass dependence, $\sigma_{\rm p2}$, $\alpha_{\rm p2}$, and $\Delta_{\rm p2}$ are consistent with zero within their 95 percent confidence intervals.  

In summary, contrary to what we found for the analysis based on the mock data, the likelihood distributions change drastically from the most restrictive model~(A) to the most flexible model~(F). Importantly, with only a few simple modifications of the CLF parametrization, the evidence increases by a staggering 63 orders of magnitude compared to the results published previously in \paperII using the same data and the same analysis but using CLF parametrization model~(B). It is worth pointing out that it is not easy to see how exactly the fit to the data has improved. Our total likelihood includes a total of 30,431 primary-secondary pairs for which we seek a model that reproduces their projected separations and line-of-sight velocities, as well as a similar number of primaries for which we use the number of secondaries as additional constraints. \Basilisk merely maximizes the likelihood for these data given a model, without actually fitting any binned data (other than the total galaxy LF, which consists of only 10 data points), which makes it difficult to assess any kind of goodness of fit by eye. Likelihood maximization of raw and unbinned data also means that \Basilisk does not produce physically interpretable $\chi^2$ values, which is why we resort to comparing everything in terms of log likelihood and Bayesian evidence.

We have performed numerous tests and find that the drastic increase in evidence, while going from model~(B) to~(E) is extremely robust. We therefore conclude that a CLF model that allows for a linear dependence of $\sigma_\rmc(M)$, $\alpha_\rms(M)$, and $\Delta_\rms(M)$, on $\log M$ is the optimum model to explain the SDSS data, at least in the context of the specific parametrizations of the CLF considered here. 

\bigskip
\bigskip

\section{Discussion}
\label{sec:discussion}

\subsection{Implications for the Galaxy-Halo Connection}
\label{sec:implications}

Fig.~\ref{fig:GHC_pars_sdss} shows the posterior predictions for the different composite parameters that make up the galaxy-halo connection in the SDSS data and how their inferences are affected by the model flexibility assumed. Panels (a) to (e) show $\Delta_\rms = \log (L_\rms^*/\bar{L}_\rmc)$, $\sigma_\rmc$, $\alpha_\rms$, $\log \bar{L}_\rmc$, and $\log \phi_s^*$, respectively, as functions of the virial mass of the host halos. We also show inferences from previous work involving a variety of techniques, including galaxy group catalogs \citep{Yang.etal.08}, a combination of galaxy clustering and weak gravitational lensing \citep{Cacciato.etal.13}, and a previous analysis of satellite kinematics \citep{Lange.etal.19b}, shown by violet stars, yellow diamonds, and blue circles, respectively. 

In panel (a), models (A-C) are fixed at $-0.25$, similar to what was done in \citet{Yang.etal.08}, \citet{Cacciato.etal.13}, and \citet{Lange.etal.19b}. However, for models (D-F), $\Delta_\rms$ is allowed to vary, with or without a mass dependence, resulting in values that are much larger, roughly $+0.2$. As we have seen above, this comes with a huge increase in the evidence, indicating that the SDSS data strongly prefer models in which the characteristic luminosity of satellite galaxies, $L_\rms^*$, is roughly 1.5 times {\it larger} than the median luminosity of central galaxies in halos of the same mass (as opposed to 1.8 times {\it smaller}, which is implied by $\Delta_\rms = -0.25$). Although this may seem unconventional, it merely reflects that, especially in massive halos, the brightest halo galaxy is not always the central one (see below for a more detailed discussion).

Panel (b) shows that model (A), which assumes that $\sigma_\rmc$ is independent of halo mass, yields a value for $\sigma_\rmc$ that is only slightly larger than the value inferred by \citet{Cacciato.etal.13}, who also assumed that $\sigma_\rmc$ is constant with halo mass.  Interestingly, if we introduce a mass dependence, i.e., going from (A) to (B), we still infer no significant mass dependence, with the results for model (B) lying virtually on top of those for model (A). This is in stark contrast to the results of \citet{Lange.etal.19b}, who used a CLF parametrization very similar to that of model (B) to also analyze satellite kinematics extracted from SDSS data, but using a method that is very different from \Basilisc. As discussed in \papII, this discrepancy is partially due to the fact that \citet{Lange.etal.19b} assumed that the central galaxy is always the brightest halo galaxy (BHG), while \Basilisk does not make this assumption. If the central is not the BHG, then the primary that we selected, and which by definition is the brightest galaxy in its selection cone, will not be the actual central. We refer to such primaries as `impurities' and \Basilisk accounts for these impurities in its analysis (see \paperII and below for details). In models (D-F), in which $\Delta_\rms$ is allowed to vary, the resultant $\sigma_\rmc(M)$ is significantly higher ($\geq 0.2$ dex) than for models (A-C). As discussed in Section~\ref{sec:intro}, different models for galaxy formation make different predictions for $\sigma_\rmc(M)$, typically with semi-analytical models predicting a scatter that is significantly larger than that predicted by hydrodynamical simulations \citep[][]{Wechsler.Tinker.18, Porras-Valverde.etal.2023}. Unfortunately, our empirical constraints on $\sigma_\rmc$ appear to strongly depend on some subtle assumptions about the flexibility of the CLF model. Interestingly, our optimal model (E) predicts values for $\sigma_\rmc(M)$ that seem to lie right between the predictions of semianalytical models and those of hydrodynamical simulations. However, we caution that a direct comparison is complicated given that the scatter inferred here is with regard to the $r$-band luminosity, while most galaxy formation models focus on the scatter in stellar mass. Finally, we emphasize that the fact that all of our results are highly discrepant with \citet{Yang.etal.08} at the low halo masses is not surprising, given that the scatter estimated from the \citet{Yang.etal.07} group catalogs is known to be systematically underestimated in low-mass halos as a consequence of the fact that the halo mass is inferred directly from the total group luminosity.  
\begin{figure*}
\centering
\includegraphics[width=0.85\textwidth]{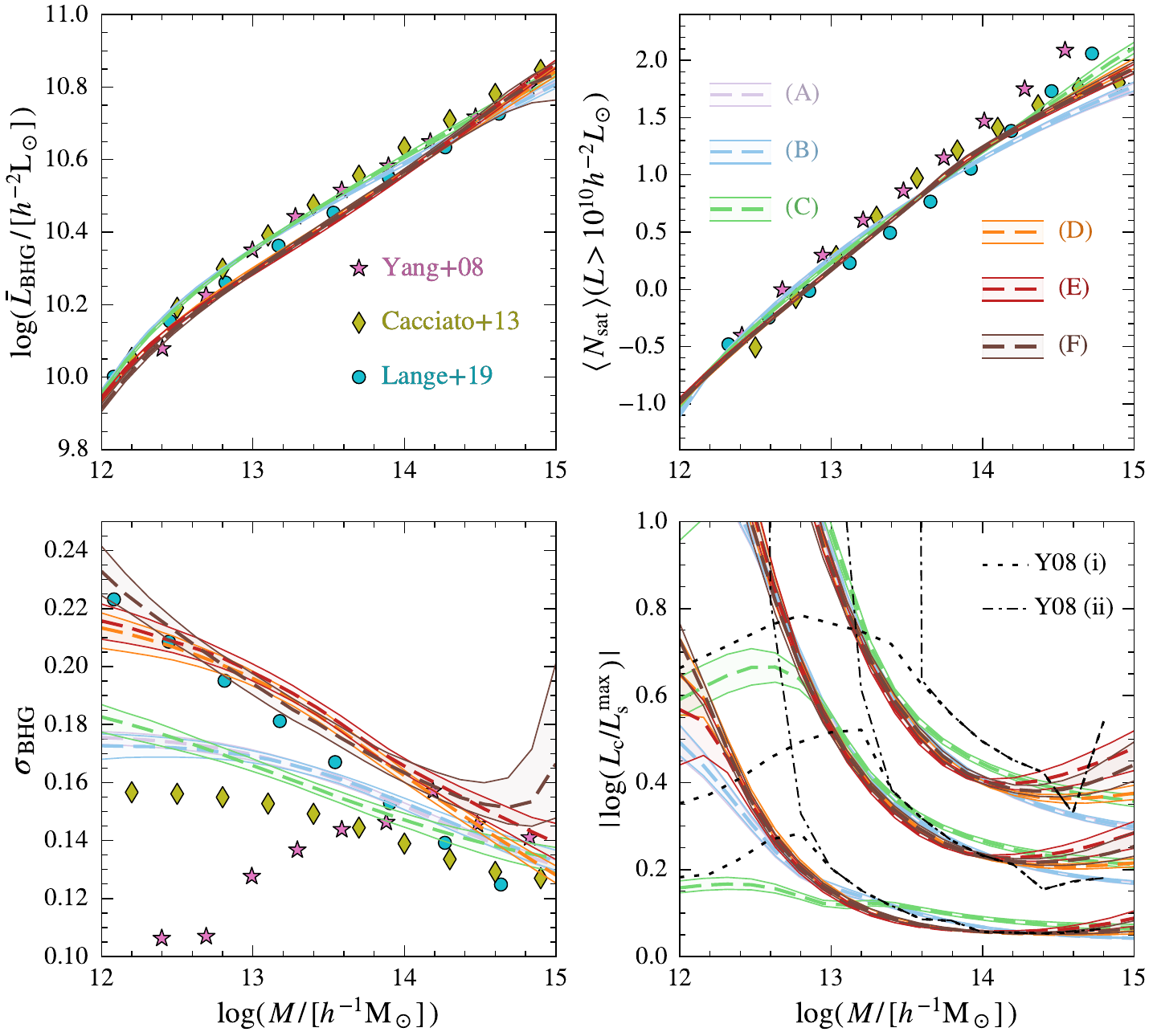}
\caption{Constraints on the halo mass dependence of properties that are more directly observable. \textit{Top-left}: the median luminosity of the brightest halo galaxy,  \textit{Top-right}: the mean number of satellite galaxies above a luminosity threshold of $10^{10} h^{-2}L_\odot$, \textit{Bottom-left}: the intrinsic scatter in the luminosities of the brightest halo galaxies. In each of these panels, the dashed lines and shaded regions of corresponding color indicate the median and 68 percentile ranges inferred from the posteriors obtained with \Basilisk analyzing the SDSS data, as indicated in the top right panel, and the symbols 
indicate constraints obtained by a variety of previous analysis of SDSS data, as indicated in the top left panel. Finally, in the \textit{Bottom-right} panel, the three sets of curves, from bottom to top, indicate the $16^{\rm th}$, $50^{\rm th}$, $84^{\rm th}$ percentiles of the distribution of the `magnitude gap', defined as $|\log (L_\rmc/L_\rms^{\rm max})|$. In this panel, the black dotted and dot-dashed curves are results from \citet{Yang.etal.08}, based on their galaxy group catalog, for two different assumptions about the occupation statistics of galaxies below the flux limit of the survey (see text for details).}
\label{fig:observables_sdss}
\end{figure*}

Panel (c) of Fig.~\ref{fig:GHC_pars_sdss} shows the faint-end slope estimates. The results for models (A) and (B) overlap exactly and are hence indistinguishable. Both models infer $\alpha_\rms \simeq -0.8$, which is in close agreement with the inference of \citet{Lange.etal.19b}, which also used satellite kinematics in the SDSS, but somewhat larger (i.e., implying a shallower faint-end slope) than the inference of \citet{Cacciato.etal.13}, which finds $\alpha_\rms \simeq -1.2$ based on an analysis of clustering and lensing in the SDSS. Note that both studies adopted a CLF model with a mass-independent $\alpha_\rms$ as in models (A) and (B). Introducing a linear mass dependence for $\alpha_\rms$, as in model (C), the posterior strongly prefers a very steep dependence on $\log M$, shown by the green dashed line,  which has $\alpha_\rms$ increasing from $-1.9$ for massive clusters to $\alpha_\rms \geq 0$ for halos with $M \lta 10^{12.2} \Msunh$. For massive halos with $M \gta 10^{13} \Msunh$, this agrees nicely with the results obtained from galaxy group catalogs by \citet{Yang.etal.08}. However, for lower mass halos the inference of model (C) implies faint-end slopes for the satellite CLF that are much shallower than what is inferred by Yang et al. As mentioned in Section~\ref{sec:modelC}, several early CLF-based studies adopted a parametrization with a mass-dependent faint-end slope. All these studies inferred a strong mass dependence, very similar to that inferred here with model (C). This includes several analyses of galaxy clustering \citep[][]{Yang.etal.03, vdBosch.etal.03, vdBosch.etal.07}, a study of the abundance and radial distribution of satellite galaxies \citep[][]{vdBosch.etal.05c}, and an analysis of the CLF inferred directly from a galaxy group catalog \citep{Yang.etal.05b}. Note that all those studies used data from the Two-Degree Field Galaxy Redshift Survey \citep[2dFGRS][]{Colless.etal.01}. Hence, there appears to be ample evidence to support a mass-dependent faint-end slope for the satellite CLF. However, the results change drastically when $\Delta_\rms$ is allowed to vary. In particular, in models (D-F) the inferred values for $\alpha_\rms$ are significantly lower, implying a much steeper faint-end slope. In fact, all three models find that $\alpha_\rms$ decreases from $\sim -1.2$ for $M = 10^{12}\Msunh$ to $\alpha_\rms = -2$ at $M \gta 10^{14} \Msunh$. Recall from Section~\ref{sec:modelC}, though, that we require $\alpha_\rms \geq -2$ to guard against models for which the total integrated satellite luminosity is infinite. Clearly, the inference of models (D-F) is impacted by this model prior.

Panel (d) of Fig.~\ref{fig:GHC_pars_sdss} shows the inferred median central luminosity as a function of host halo mass. The results of models (A-C) overlap each other and are in excellent agreement with the results of \citet{Yang.etal.08}, \citet{Cacciato.etal.13}, and \citet{Lange.etal.19b}. However, for models (D-F), in which $\Delta_\rms$ is allowed to vary, the inferred $\bar{L}_\rmc(M)$ is significantly lower. A similar trend is apparent for the normalization of the satellite CLF, $\phi_\rms^*$, shown in panel (e). We address these discrepancies in detail below.

%\subsection{The galaxy-halo connection of brightest halo galaxies: A much fairer comparison with past results}
\subsection{Halo occupation statistics of Brightest Halo Galaxies: a more direct comparison
with observations}

Given that models (D-F) are strongly preferred by the data over models (A-C), at face value these results suggest that the galaxy-halo connection inference from many prior studies might be systematically biased. However, it is important to realize that it is typically assumed that the central galaxy is synonymous with the BHG. For instance, \citet{Yang.etal.08} assign the brightest group member the status of central galaxy, while \citet{Lange.etal.19b} use an isolation criterion similar to that used here, in which primaries are defined as the brightest galaxies in their selection cone. In their subsequent analysis of satellite kinematics, they implicitly assumed that the primary is always the central. Although \Basilisk also requires primaries to be the brightest galaxies in their selection cone, it accounts for the possibility that the primary is not the central in its computation of the probability $P(L_{\rm pri}|M)$. If primaries are always centrals, then this probability function is simply given by the central CLF, that is, $P(L_{\rm pri}|M) = \Phi_\rmc(L_{\rm pri}|M)$. However, in \Basilisk we use
\begin{align}\label{Pbhg}
P(L_{\rm pri}|M) = & P(L_\rmc = L_{\rm pri}|M) \,\, P(L_{\rm bs} < L_{\rm pri}|M) \nonumber \\
& P(L_\rmc < L_{\rm pri}|M) \,\, P(L_{\rm bs} = L_{\rm pri}|M)
\end{align}
where $L_{\rm bs}$ indicates the luminosity of the brightest satellite. As shown in \papII, these probabilities can be calculated directly from the CLF, and using equation~(\ref{Pbhg}) successfully mitigates potential biases in inference caused by impurities (i.e., primaries that are not centrals).

Under the assumption that satellite galaxies follow Poisson statistics, the probability that a halo of mass $M$ hosts zero satellites brighter than $L$ is given by $\exp[-\Lambda(L)]$, where 
\begin{equation}\label{Lambdasat}
\Lambda(L) = \int_L^\infty \Phi_\rms(L'|M) \, \rmd L'
\end{equation}
is the expectation value for the number of such satellites. Hence, the probability that a halo of mass $M$ has a central of luminosity $L$ and at least one satellite brighter than the central is given by $\Phi_\rmc(L|M) \left(1-\exp[-\Lambda(L)] \right)$. Integrating over all possible luminosities for the central then yields the probability $P_{\rm impurity}(M)$ that in a halo of mass $M$ the central is not the BHG, i.e.,
\begin{equation}
P_{\rm impurity}(M) = \int_0^{\infty}\Phi_\rmc(L'|M) \, \left(1-\exp[-\Lambda(L')] \right) \, \rmd L'\,.
\end{equation}

Fig.~\ref{fig:cen_bhg} shows the posterior predictions for $P_{\rm impurity}(M)$ of models (A-F). This highlights the most important difference between models (A-C) on the one hand, and models (D-F) on the other hand; the latter predict much larger impurity fractions at the high halo mass end. For comparison, the various symbols indicate observational constraints from a variety of different techniques, as indicated. Although these seem to be in better agreement with models (D-F) than with models (A-C), the observational uncertainties are large. For instance, the impurity is defined in terms of luminosity, computed from the SDSS r-band Petrosian magnitude. As discussed in Section~\ref{sec:concl}, these have substantial uncertainties that can impact the inferred impurity fractions, especially for brightest cluster galaxies. More general, though, it is difficult to unambiguously identify the central galaxy in large groups and clusters of galaxies, especially in a redshift survey without auxiliary X-ray data. 

However, it is quite trivial to identify the BHG. Therefore, in Fig.~\ref{fig:observables_sdss} we show the posterior predictions of models (A-F) in terms of the statistics of the BHGs rather than in terms of centrals and satellites. The top left panel shows the relation between the median $L_{\rm bhg}$ and halo mass, calculated from the CLF using 
\begin{equation} \label{Lbhg}
\bar{L}_{\rm BHG}(M) = \int \, L_{\rm pri} \, P(L_{\rm pri}|M) \, \rmd L_{\rm pri}
\end{equation}
where $P(L_{\rm pri}|M)$ is given by equation~(\ref{Pbhg}). The symbols show the same for the analyzes of \citet{Yang.etal.08}, \citet{Cacciato.etal.13}, and \citep{Lange.etal.19b}. In the case of Yang et al., since they {\it define} centrals as the brightest group members, this is identical to the results shown in Fig.~\ref{fig:GHC_pars_sdss}. Similarly, Lange et al. did not make any distinction between the centrals and the primaries, which are almost always the BHGs, and thus the cyan circles are also the same as those in Fig.~\ref{fig:GHC_pars_sdss}. In the case of Cacciato et al., we used their best-fit CLF models to calculate the median luminosity of the BHG using equation~(\ref{Lbhg}). Note that the mutual agreement here is much better than in panel (d) of Fig.~\ref{fig:GHC_pars_sdss}. 

A similar level of consistency is evident in the top right panel of Fig.~\ref{fig:observables_sdss}, which shows the total number of satellite galaxies brighter than $L_{\rm th} \equiv 10^{10} \Lsunh$, calculated using
\begin{equation}
\langle N_{\rm sat} \rangle = \int_{L_{\rm th}} \Phi_{\rm s}(L'|M) \, \rmd L'\,.
\end{equation}
Note that all six CLF models are in excellent agreement with each other, and with the constraints from previous analysis. Note that the results of \citet{Yang.etal.08} suggest a larger number of satellites in massive halos, but we suspect that these results are biased by interlopers (galaxies in the fore or background of the galaxy groups). The fact that all models agree on the observable $\langle N_{\rm sat} \rangle$, elucidates that the discrepancy in the predictions for $\Phi_\rms^*(M)$ shown in panel (e) of Fig.~\ref{fig:GHC_pars_sdss} is of no concern. What is relevant are the predictions for observables, not the parametrization used. In particular, in the case of our CLF, the observable $\langle N_{\rm sat} \rangle$ is given by $\phi_\rms^* \Gamma([1+\alpha_\rms]/2, [L_{\rm th}/L_\rms^*]^2)$, with $\Gamma(a,x)$ the incomplete Gamma function. Hence, keeping $\langle N_{\rm sat} \rangle$ invariant requires $\phi_\rms^*$ to compensate for any changes in $\alpha_\rms$ and/or $L_\rms^*$.
\begin{figure*}
\centering
\includegraphics[width=0.85\textwidth]{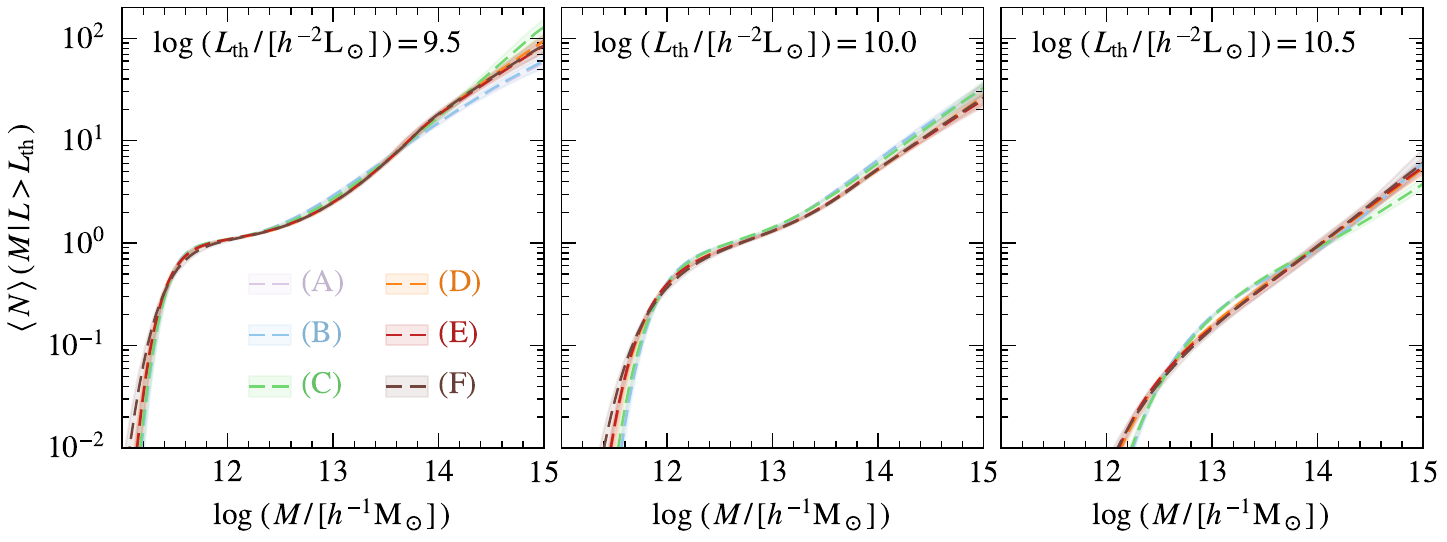}
\caption{The halo occupation distributions (HODs) predicted from \Basilisc's posterior constraints on the CLF as inferred from the SDSS data. Different panels show the HODs for three different luminosity thresholds, as indicated. Different sets of colored lines are the predicted HODs for the six different CLF models (A-F), as indicated in the left panel. Note that these six models predict HODs that are in perfect agreement with each other, despite the fact that they predict CLFs that are starkly different. It is also noteworthy that the shapes of these HODs look remarkably similar to the standard functional form \citep{Zheng.etal.07} typically adopted in the literature, even though there is no a priori reason that this would be the case.}
\label{fig:HOD_sdss}
\end{figure*}

The bottom-left panel of Fig.~\ref{fig:observables_sdss} shows the scatter in the luminosities of the brightest halo galaxies, rather than the central galaxies. This is calculated using
\begin{equation} \label{sigbhg}
\sigma^2_{\rm BHG}(M) = \int \, (\log L_{\rm pri}-\log\bar{L}_{\rm BHG})^2 \, P(L_{\rm pri}|M) \, \rmd L_{\rm pri} \,,
\end{equation}
For comparison, we also show $\sigma_\rmc$ from the same previous analyses as before. In the cases of Yang et al. and Lange et al., we adopt $\sigma_{\rm BHG} = \sigma_\rmc$, which is reasonable given their assumption that true centrals always are the BHGs. For \citet{Cacciato.etal.13}, we instead compute $\sigma_{\rm BHG}$ from their best-fit CLF model using equation~(\ref{sigbhg}). Recall that the \citet{Yang.etal.08} scatter is known to be systematically underestimated in low-mass groups, which explains why their results deviate significantly from our predictions and other previous results for $M \lta 10^{13.5} \Msunh$. Note that $\sigma_{\rm bhg}$ inferred from our analysis with \Basilisk depends strongly on whether $\Delta_\rms$ is kept free (models A-C), or is allowed to vary (models D-F). Crucially, though, all models predict very similar $\sigma_{\rm bhg}$ at the massive end, and all models agree that the scatter increases with decreasing halo mass. This is very different from the predictions for $\sigma_\rmc(M)$ shown in panel (b) of Fig.~\ref{fig:GHC_pars_sdss}. This is reassuring, since $\sigma_{\rm BHG}$ is not affected by ambiguity in deciding which galaxy is the central one and is thus more directly observable.
%Interestingly, the more flexible models agree very well with \citet{Lange.etal.19b}.

Finally, another observable quantity of physical interest is the magnitude-gap distribution in galaxy groups and clusters. This is defined as the magnitude difference, or equivalently, the logarithm of the luminosity ratio, between the two brightest galaxies in the group or cluster. If we assume that the two brightest galaxies are always the central and the brightest satellite (i.e., we assume that the probability that the second brightest satellite outshines the central is negligible), then we can compute this ratio as $|\log ( L_\rmc / L_\rms^{\rm max})| $, where $L_\rms^{\rm max}$ is the luminosity of the brightest satellite. Note that by taking the absolute value we are agnostic as to whether the BHG is the central or the brightest satellite. 

The luminosity of the brightest satellite, $L_\rms^{\rm max}$, follows a probability distribution $P_{\rm s,max}$ that can be computed by combining the probability for a satellite to have luminosity $L_\rms^{\rm max}$, which follows from the satellite CLF, and the probability that there are no satellites with $L > L_\rms^{\rm max}$. Assuming that satellite galaxies follow Poisson statistics, this implies that
\begin{equation} \label{eqn:P_Lsmax2}
P_{\rm s,max}(L) = \dfrac{\Phi_\rms(L) \, \exp\left[ - \Lambda(L) \right]}{\int \rmd L' \,\, \Phi_\rms(L') \, \exp\left[-\Lambda(L') \right]}\,,
\end{equation}
with $\Lambda(L)$ the expected number of satellites brighter than $L$ given by equation~(\ref{Lambdasat}).  If we define the magnitude gap as $\Delta \log L =  |\log L_\rmc - \log L_\rms^{\rm max}|$, and assume that the CLFs of centrals and satellites are independent\footnote{Although this is a standard assumption throughout, it is questionable given that central galaxies can cannibalize satellites.}, then the probability distribution of $\Delta \log L$ is given by
\begin{equation}
    P_{\rm gap}(\Delta \log L) = \int \rmd L_\rmc \,\Phi_\rmc(L_\rmc) \, \left[ P_{\rm s,max}(L_{-}) +  P_{\rm s,max}(L_{+})\right]\,. 
\end{equation}
Here, $\log L_{\pm} = \log L_\rmc \pm \Delta \log L$, representing the two cases where the central is brighter and fainter than the brightest satellites. 

For each of our models (A-F), the color-shaded curves in the bottom-right panel of Fig.~\ref{fig:observables_sdss} show, from top to bottom, the $84^{\rm th}$, $50^{\rm th}$, and $16^{\rm th}$ percentiles of $P_{\rm gap}(\Delta \log L)$. All six models are in close agreement, with the exception of model (C) which deviates from the others at the low mass end. All models predict that $\Delta\log L$ increases strongly with decreasing halo mass but only for $M \lta 10^{13.5} \Msunh$. Above that mass, all models agree that the median expectation value for $\Delta \log L \simeq 0.25$.  For comparison, the black dotted and dot-dashed lines show the corresponding percentiles of $P_{\rm gap}(\Delta \log L)$ obtained by \citet[][hereafter Y08]{Yang.etal.08} using their SDSS group catalog (cf. their Fig.~7). Note that we show two different predictions for these percentiles, depending on the assumption made about the luminosity of the second brightest galaxy in groups with only one member. The dotted curves labeled Y08 (i) are the results obtained if it is assumed that the second brightest galaxy in these ``lonely'' groups has a luminosity just barely below the survey flux limit. The dot-dashed curves labeled Y08 (ii) correspond to the alternative extreme choice, where it is assumed that these groups are truly lonely and do not have a second satellite, no matter how faint. This is equivalent to assuming that the luminosity of the second-brightest galaxy is zero. The true magnitude gap percentiles are expected to lie somewhere between these two extreme, limiting cases. Note that \Basilisc's constraints on the percentiles as a function of halo mass are in reasonable agreement with those of Y08. Although the agreement is very good for the $50^{\rm th}$ percentiles, the Y08 results generally predict $16^{\rm th}$ and $84^{\rm th}$ percentiles that are somewhat larger than what we infer with \Basilisc. Nevertheless,
the agreement is satisfactory, both qualitatively and quantitatively. 

In summary, as long as we focus on observables that are agnostic as to whether the central is the BHG or not, all six models are in much better agreement with each other and also with previous constraints. 

\subsection{The Halo Occupation Distribution}
\label{sec:hod}

So far we have only discussed the galaxy-halo connection in terms of the CLF. However, in the literature, it is more common to characterize the statistical link between galaxies and their host halos in terms of the halo occupation distribution,  which characterizes the expectation value for the number of galaxies (centrals plus satellites) brighter than some specified luminosity threshold, $L_{\rm th}$, as a function of halo mass. We can trivially compute the HODs for our various models from the inferred CLF using
\begin{equation}
 \langle N \rangle(M) = \int_{L_{\rm th}}^{\infty} \Phi(L|M) \, \rmd L\,.
\end{equation}
Fig.~\ref{fig:HOD_sdss} shows the predicted HODs, for models (A-F) for three different values of $L_{\rm th}$ as indicated. Remarkably, despite the fact that the CLFs for models (A) to (F) vary drastically, as demonstrated above, their corresponding HOD predictions are almost indistinguishable. Hence, it appears that \Basilisk is able to put extremely tight and robust constraints on the occupation {\it numbers} of galaxies. However, when it comes down to assigning individual luminosities to these galaxies, the inference strongly depends on the CLF parametrization adopted. On the one hand, this should not come as a huge surprise, given that the CLF contains more information than the HOD, but it is nonetheless surprising to see how extremely consistent the HOD inference is.

Not only is our HOD inference robust, it also has a shape that is remarkably consistent with standard assumptions about the HOD, which consists of an error function that describes the occupation statistics of centrals and a power law to describe the occupation statistics of the satellites \citep{Zheng.etal.05, Zheng.etal.07}. In particular, it is surprising to see how well the satellite HOD is approximated by a simple power law, even though such a form is by no means guaranteed given the functional forms we have adopted for the CLF. All of this suggests that the standard HOD model that is often adopted seems to be empirically supported. 

\section{Summary and Conclusion}
\label{sec:concl}

In this work, we used observational data on the abundance and kinematics of satellite galaxies to explore how assumptions about the functional form used to characterize halo occupation statistics impact the inference on the galaxy-halo connection. Specifically, we used \Basilisc, a powerful Bayesian hierarchical tool for forward modeling the kinematics and abundance of satellite galaxies, to test the standard CLF model against several more flexible variants. This includes several degrees of freedom that have previously been unexplored.

We first tested our approach using realistic SDSS-like mock data, for which the underlying galaxy-halo connection is known precisely. Using \Basilisk we re-analyzed these mock data multiple times, each time using a different CLF with more and more model freedom, going from the most basic CLF model with 9 free parameters, which is the one used to construct the mock data, to the most flexible model with 16 free parameters. We use the Bayes factor to compare the different models among themselves, which properly penalizes an increase in model freedom that does not result in a significant improvement in the likelihood. We infer that the optimal model, defined as the model with the largest Bayes factor, is indeed the most basic model. This indicates that the excessive freedom in the more extensive models does not introduce any bias in our inference. In fact, for all model variations we correctly recover the true underlying galaxy-halo connection, irrespective of the flexibility assumed in the CLF. As expected, using a more flexible model yields posteriors that are broader (that is, weaker constraints on the parameters), but the posteriors remain consistent with the input values for all model variations used. Hence, based on this mock validation, we conclude that \Basilisk is able to find the optimal model required by the data, without unnecessary degrees of freedom resulting in a biased inference.

Next, we applied \Basilisk to SDSS DR-7 data using the same set of CLF models with different degrees of flexibility. Most of these models are significantly more flexible than the `standard model' used in our previous analysis of \citep[][]{Mitra.etal.24}, which was based on exactly the same data. One model clearly stands out as the optimal, beyond with added model flexibility does not improve the fit to the data and is therefore disfavoured in terms of the Bayes factor. This optimum model differs from the `standard' CLF parametrization in the literature, which is also the one used in \citet{Mitra.etal.24}, in terms of 3 model extensions: (i) halo mass dependence in the scatter of the luminosities of centrals, (ii) halo mass dependence in the faint-end slope of the satellite CLF, and (iii) allowing halo mass dependent freedom in the characteristic luminosity of satellite galaxies such that it is no longer restricted to a fixed fraction of the characteristic luminosity of its central. The Bayes factor of this optimal model is larger than that of the `standard model' used in \citet{Mitra.etal.24} by a staggering factor of $10^{63}$, indicating that the SDSS data strongly demands the extra degrees of freedom provided by the optimal model.

When using the inflexible standard model in our analysis, key aspects of the inferred galaxy-halo connection, such as the relation between halo mass and (median) central galaxy luminosity, its scatter, and the normalization of the satellite CLF, are all in excellent agreement with past results based on galaxy group catalogs, galaxy clustering, galaxy-galaxy lensing, and previous analyzes of satellite kinematics. However, \Basilisc's constraints based on the optimal model deviate significantly from these `established' results. However, we have shown that this apparent discrepancy is mainly due to issues related to misidentifying the brightest halo galaxy (BHG) as the central galaxy. In particular, we showed that the posteriors obtained with \Basilisk for all CLF model extensions are in reasonable agreement with each other and with the prior results when it comes to the halo occupation statistics of BHGs. Observationally, it is more direct to identify the BHG of a halo rather than its central, which  suggests that the occupation statistics of the former are a more reliable constraint for the models.

Fortunately, it is straightforward to compute these BHG statistics from the CLF. We use this to compute, for each of the different parametrizations of the CLF model used, the inferred central galaxy impurity fraction, defined as the fraction of halos of a given mass in which the central galaxy is {\it not} the BHG. In the halo mass range where we have sufficient constraining power ($10^{12.5}< M < 10^{14.5} \Msunh$), the predictions from our more flexible models, including the optimal one, agree very well with previous results based on a variety of different probes. However, the less flexible models, including the standard one, predict impurity fractions that are too low, especially in massive halos. According to \Basilisc's inference based on the optimal CLF model, the impurity fraction of SDSS galaxies increases from close to zero for $M < 10^{12} \Msunh$ to about 50 percent for $M=10^{14}\Msunh$, and as high as 80 percent for $M=10^{15}\Msunh$. These large fractions may seem surprising given that brightest cluster galaxies (BCGs) are generally considered the most massive galaxies known \citep[e.g.,][]{Lin.Mohr.04, Lauer.etal.14}. We emphasize, though, that our inference of the impurity fraction is based on the Petrosian $r$-band luminosities of galaxies in the SDSS. It is well known that these carry significant uncertainties due to issues with deblending, sky background subtraction, and the contribution of a stellar halo or intra-cluster light. As several studies have pointed out, these issues are especially acute for BCGs, and are likely to cause their Petrosian SDSS magnitudes to be severely underestimated \citet{Bernardi.etal.10, Bernardi.etal.13, Simard.etal.11, Souza.etal.15}. Hence, our inference that in the vast majority of massive clusters the central galaxy is not the brightest halo galaxy, has to be taken with a grain of salt in light of these uncertainties.

Despite the large variation in the CLF constraints among the different model parametrizations, the inferred HODs, which quantify the average halo occupation numbers of galaxies above some given magnitude limit, are almost exactly overlapping. Moreover, the shapes of the inferred HODs are remarkably consistent with the standard functional form adopted in the literature, despite the fact that the CLF models have sufficient flexibility to deviate significantly from it. This indicates that \Basilisk can put extremely tight and robust constraints on the {\it numbers} of galaxies as a function of halo mass. However, when it comes to assigning luminosities to these galaxies, and deciding on which one is the true central, many uncertainties enter the analysis. While this seems to be an argument against using the CLF, and in favor of using the much simpler HOD models, we emphasize that CLF constraints are far more powerful for constraining galaxy formation models and  thus should not be abandoned. 

To conclude, the analysis presented here highlights that any model parametrization, be it an empirical model for galaxy formation or a model characterizing halo occupation statistics, acts as an implicit prior that can significantly bias the inference. We have demonstrated the importance of adding degrees of freedom to the model until this is no longer warranted by the data. Unfortunately, such an approach can never be fully exhaustive. In particular, the exploration of CLF model freedom explored here and the fact that an optimal model was identified do not preclude the possibility that the true CLF has a functional form that is not captured by any of the forms considered here. Testing such an hypothesis will likely require a non-parametric form for the CLF which may be within reach in the next decade with new data from the Dark Energy Spectroscopic Instrument \citep[DESI;][]{Desi.22} Bright Galaxy Survey \citep[BGS;][]{DESI.BGS.23}. Hence, there is ample reason to be optimistic that with better data and improved modelling, we will continue to tighten the constraints on the galaxy-dark matter connection, thereby informing both our models of galaxy formation and our cosmological framework. As a specific example, in a follow-up paper in this series \citep[][]{Mitra.etal.25}, we use \Basilisk in combination with the optimal CLF model identified here, to infer constraints on cosmological parameters that are independent from most other studies that rely mainly on large-scale data. 

\section*{Acknowledgements}

This work has been supported by the National Aeronautics and Space Administration through Grant No. 19-ATP19-0059 issued as part of the Astrophysics Theory Program. FvdB has received additional support from the National Science Foundation (NSF) through grants AST-2307280 and AST-2407063. This work utilized, primarily for plotting purposes, the following python packages: \texttt{Matplotlib} \citep{Matplotlib_Hunter2007}, \texttt{SciPy} \citep{SciPy_Virtanen2020}, \texttt{NumPy} \citep{numpy_vanderWalt2011}, and \texttt{PyGTC} \citep{PyGTC_Bocquet2016}.

%%%%%%%%%%%%%%%%%
% Bibliography
%%%%%%%%%%%%%%%%%

\bibliographystyle{mnras}
\bibliography{references_vdb}

%%%%%%%%%%%%%%%%%%%%%%%%%%%%%%%%%%%%%%%%%%%%%%%%%%%%%%%%%%%%%%%%%%%%%%%%%%

\appendix
\numberwithin{figure}{section}
\numberwithin{table}{section}
\numberwithin{equation}{section}

\section{Computing Bayesian Evidence}
\label{App:evidence_calculation}

In Bayesian parameter estimation, the posterior probability is given by
\begin{equation}
    P(\btheta|\bD) = \dfrac{\calL(\bD|\btheta) \, P(\btheta)}{\calZ(\bD)}\,,
\end{equation}
where $\calL(\bD|\btheta)$ is the likelihood for the data $\bD$ given the parameters $\btheta$ and $P(\btheta)$ is the corresponding prior distribution.  The Bayesian evidence, $\calZ(\bD)$, plays little to no role in the parameter estimation process, but is crucial when it comes to model comparison: to decide which model better explains the data. If we consider two models, $\calM_1$ and $\calM_2$, with their own set of parameters ($\btheta_1$ and $\btheta_2$, respectively), the plausibility of one model ($\calM_1$) compared to another ($\calM_2$) is evaluated in terms of the Bayes factor:
\begin{equation} \label{eqn:Bayes_factor_1}
    \calK = \dfrac{\calZ(\bD|\calM_1)}{\calZ(\bD|\calM_2)} = \dfrac{\int {\rm d} \btheta_1 \, \calL(\bD|\btheta_1,\calM_1) \, P(\btheta_1|\calM_1)}{\int {\rm d} \btheta_2 \, \calL(\bD|\btheta_2,\calM_2) \, P(\btheta_2|\calM_2)} \,.
\end{equation}
This integral is not trivial to compute directly for our models with the number of free parameters ranging between 13 and 20. Nested sampling \citep{Buchner.2023} is a common approach to tackle similar issues. However, \Basilisk in its current form uses an affine invariant MCMC for the inference procedure. Therefore, we resort to the following technique, which is valid only for cases where (i) the priors are flat or uninformed, and (ii) the posterior distribution is approximately a uni-modal multivariate Gaussian. In our case, these conditions are satisfied (see \papII).

For each of the models explored in this work, all parameters have broad, flat, uninformative priors. Thus, for any model $i$, the prior is $P(\btheta_i|\calM_i) = V_i^{-1}$, where the allowed parameter space volume is $V_i = \int {\rm d} \btheta_i$. Thus, the evidence can be written as $\calZ(\bD|\calM_i) = V_i^{-1} \int {\rm d}\btheta_i \, \calL(\bD|\btheta_i, \calM_i)$. If the number of parameters in the model $\calM_i$ is $N_i$, then the evaluation of $\calZ(\bD|\calM_i)$ becomes an $N_i$-dimensional integral. It can be transformed into a one-dimensional integral by first identifying iso-likelihood contours in the parameter space and then applying the principle of Lebesgue integration across those iso-likelihood shells. We do that by using the posterior distribution of converged MCMC walkers in the parameter space. This works because the iso-density contours of the walkers are proxies for the iso-likelihood contours.
\begin{figure}
\centering
\includegraphics[width=0.5\textwidth]{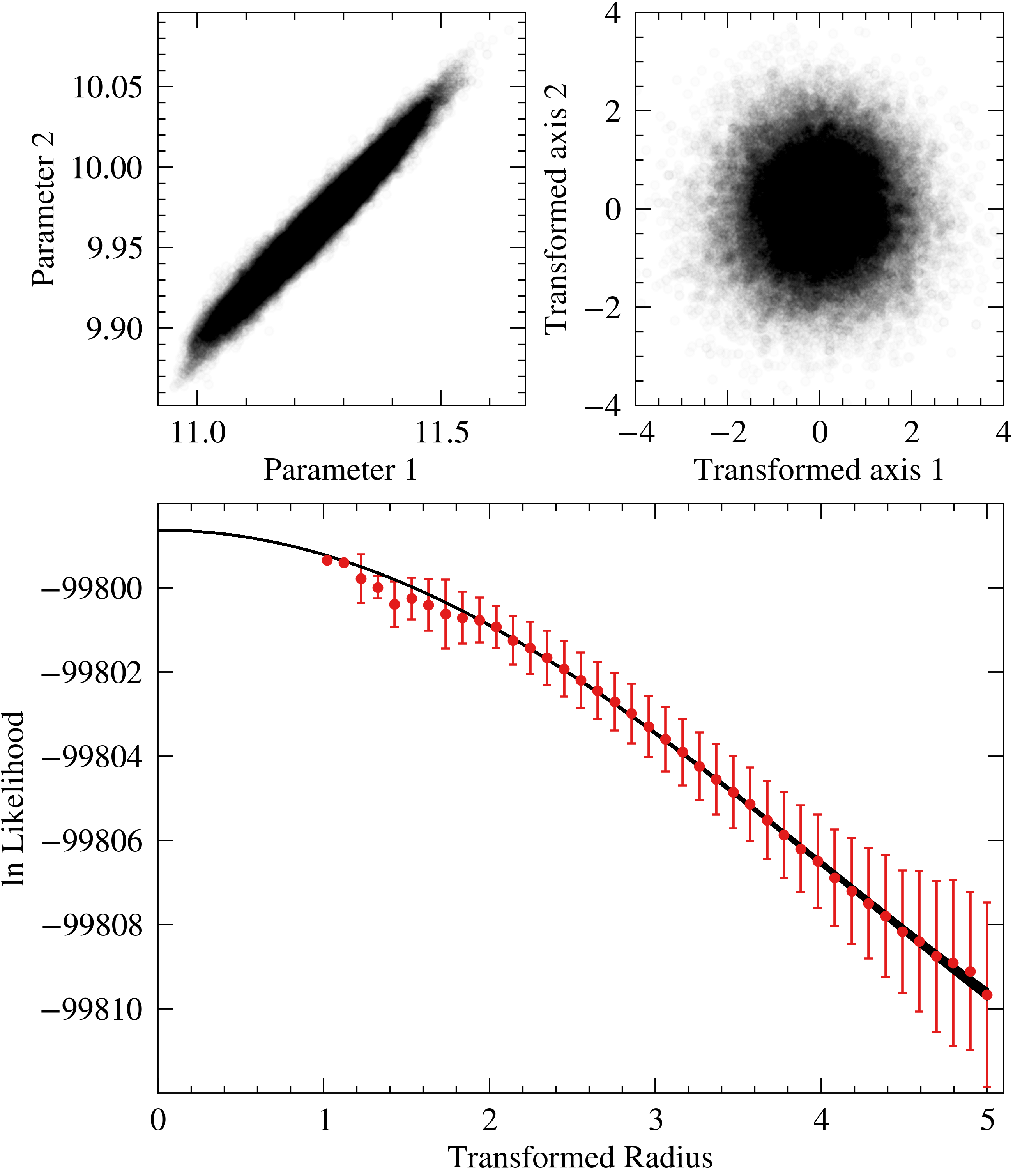}
\caption{Illustration of the method used to computing the evidence using a linear transformation of the posterior distribution of parameters. \textit{Top-left}: The posterior distribution of two arbitrarily chosen parameters (specifically $\log M_1$ and $\log L_0$) from the MCMC chain obtained fitting the mock data as described in Section~\ref{sec:mock_validation}. \textit{Top-right}: The same posterior distribution, but this time shown along two arbitrarily chosen eigen-axes in the linearly transformed space, showing a random projection of a $N$-dimensional hyper-sphere. \textit{Bottom}: The points with error-bars show the mean and standard deviations of the log likelihoods in radial bins (or hyper-shells) in the linearly transformed space. The black shaded region (almost a line) is the 95\% credible interval of a quadratic fit to the points. The Bayesian evidence is calculated by integrating this fitting function of $\ln \calL(r)$ over the hyper-sphere in the transformed space .}
\label{fig:Evidence}
\end{figure}

Specifically, we use singular value decomposition\footnote{https://scikit-learn.org/stable/api/sklearn.decomposition.html} to alter the posterior distribution of MCMC walkers in the original parameter space $\{ \theta_i \}$ through a series of linear transformations (rotation and rescaling) to reach a new distribution of points in the transformed space $\{ \tilde{\theta}_i \}$. The rescaling is such that all points on a 1$\sigma$ hyper-ellipsoid in the $\{ \theta_i \}$ space now lie on the hypersurface of a unit $N_i$-sphere in the $\{ \tilde{\theta}_i \}$ space. This linear transformation of the posterior parameter distribution, from the $\{ \theta_i \}$-space to the $\{ \tilde{\theta}_i \}$-space, is shown in the top two panels of Fig.~\ref{fig:Evidence}. The top-left panel shows the projection of the posterior distribution in terms of two arbitrarily chosen parameters for the MCMC chain resulting from \Basilisk fitting the Mock survey data with model (A). The top-right panel shows a random projection of the transformed distribution of points in the corresponding $\{ \tilde{\theta}_i \}$-space.

Each radial shell in the $\{ \tilde{\theta}_i \}$ space is approximately an iso-likelihood hypersurface. We fit a quadratic functional form to the logarithmic likelihood ($\ln \calL$) as a function of the radial distance in the transformed space, shown in the bottom panel of Fig.~\ref{fig:Evidence}. Using this fitting function ($\ln \tilde{\calL}(r)$), we can trivially compute the integral
\begin{equation}
 \int \rmd\btheta_i \, \calL(\bD|\btheta_i,\calM_i) = \int \rmd \tilde{\btheta}_i \, \calL(\bD|\tilde{\btheta}_i,\calM_i) \,\, \left( \dfrac{\rmd\theta_i}{\rmd\tilde{\theta}_i} \right) = \dfrac{2 \pi^{N_i/2}}{\Gamma(N_i/2)} \int \rmd r \, r^{N_i-1} \, \exp \left[\ln \tilde{\calL}(r) \right] \,\, \left( \prod_{k=1}^{N_i} \lambda_{i,k} \right) \,.
\end{equation}
The first equality refers to the transformation from the $\{\theta_i\}$- to the $\{\tilde{\theta}_i\}$-space, where the Jacobian (${\rm d}\theta_i / {\rm d}\tilde{\theta}_i$) captures how the volume element transforms. The second equality converts the $N_i$-dimensional integral to a 1-D integral over radial shells. Each shell has a differential volume related to the area of the hypersurface of an $N_i$-sphere, given by $\left[ 2 \pi^{N_i/2} / \Gamma(N_i/2) \right] \, r^{N_i-1} \, \rmd r$. The volume element ratio between the two complementary spaces, given by the Jacobian (${\rm d}\theta_i / {\rm d}\tilde{\theta}_i$), is the same as the ratio of the volume of a hyper-ellipsoid with the axes lengths given by the eigenvalues of the transformation matrix $\lambda_{i,k}$ (where $k = 1,N_i$) and the volume of an $N_i$-dimensional unit-sphere. This is simply the product of the eigenvalues, $\prod_k \lambda_{i,k}$, which substitutes the Jacobian in the last equality. We have repeatedly tested and ensured that, due to the sharp exponential drop in likelihood, truncating the integral to a maximum radius $r_{\rm max}$ of any value greater than $r_{\rm max}=4$ makes no difference to the value of the integral. Therefore, we adopt a truncation radius of $r_{\rm max}=5$ throughout.

For broad uninformed priors in the $i^{\rm th}$ model, given by $\{\theta_{i,k}^{\rm max},\theta_{i,k}^{\rm min}\}$, where $k$ ranges from $1$ to $N_i$, the total volume allowed for Markovian walkers is $V_i = \prod_{k=1}^{N_i} (\theta_{i,k}^{\rm max}-\theta_{i,k}^{\rm min})$. Thus, combining this with the integral over the likelihood, we can rewrite equation~\ref{eqn:Bayes_factor_1}, the Bayes factor for model $\calM_1$ compared to model $\calM_2$, as:
\begin{equation} \label{eqn:Bayes_factor_2}
  \calK = \dfrac{\pi^{N_1/2} \, \Gamma(N_2/2) \, \prod_{k=1}^{N_2} (\theta_{2,k}^{\rm max} - \theta_{2,k}^{\rm min}) \int \rmd r \, r^{N_1-1} \, \exp\left[\ln\tilde{\calL}_1(r) \right] \, \left( \prod_{k=1}^{N_1} \lambda_{1,k} \right)} {\pi^{N_2/2} \, \Gamma(N_1/2) \, \prod_{k=1}^{N_1} (\theta_{1,k}^{\rm max} - \theta_{1,k}^{\rm min}) \int \rmd r \, r^{N_2-1} \, \exp\left[\ln\tilde{\calL}_2(r) \right] \, \left( \prod_{k=1}^{N_2} \lambda_{2,k} \right)}\,.
\end{equation}
The prefactors outside the integrals are known for any given model and its predefined priors. Inside the integrals, the eigenvalues ($\lambda_{i,k}$) are obtained from the singular value decomposition (top panels of Fig.~\ref{fig:Evidence}), and $[\ln\tilde{\calL}_i(r)]$ is obtained from the fit shown in the bottom panel of Fig.~\ref{fig:Evidence}.

\label{lastpage}
\end{document}